\documentclass[12pt,authoryear]{elsarticle}

\usepackage{graphicx}
\usepackage{amssymb}
\usepackage{amsthm}
\usepackage{lineno}

\biboptions{comma,round,semicolon}

\journal{Icarus}

\begin{document}

\begin{frontmatter}


\title{The Reflectivity of Mars at 1064 nm: Derivation from Mars Orbiter Laser Altimeter Data and Application to Climatology and Meteorology}



\author[label1]{N.G. Heavens}

\address[label1]{Department of Atmospheric and Planetary Sciences, Hampton University, 23 E. Tyler St., Hampton, Virginia, 23669.}
\ead{nicholas.heavens@hamptonu.edu}

\begin{abstract}
The Mars Orbiter Laser Altimeter (MOLA) on board Mars Global Surveyor (MGS) made $> 10^{8}$ measurements of the reflectivity of Mars at 1064 nm ($R_{1064}$) by both active sounding and passive radiometry. Past studies of $R_{1064}$ neglected the effects of atmospheric opacity and viewing geometry on both active and passive measurements and also identified a potential calibration issue with passive radiometry. Therefore, as yet, there exists no acceptable reference $R_{1064}$ to derive a column opacity product for atmospheric studies and planning future orbital lidar observations. Here, such a reference $R_{1064}$ is derived by seeking $R^{M,N}_{1064}$: a Minnaert-corrected normal albedo under clear conditions and assuming minimal phase angle dependence. Over darker surfaces, $R^{M,N}_{1064}$ and the absolute level of atmospheric opacity were estimated from active sounding. Over all surfaces, the opacity derived from active sounding was used to exclude passive radiometry measurements made under opaque conditions and estimate $R^{M,N}_{1064}$. These latter estimates then were re-calibrated by comparison with $R^{M,N}$ derived from Hubble Space Telescope (HST) observations over areas of approximately uniform reflectivity. Estimates of $R^{M,N}_{1064}$ from re-calibrated passive radiometry typically agree with HST observations within 10 \%. The resulting $R^{M,N}_{1064}$ is then used to derive and quantify the uncertainties of a column opacity product, which can be applied to meteorological and climatological studies of Mars, particularly to detect and measure mesoscale cloud/aerosol structures.                 

\end{abstract}


\end{frontmatter}


\section{Introduction}
\label{S:1}
The Mars Orbiter Laser Altimeter (MOLA) was primarily designed to measure the topography of Mars at high enough precision and resolution to answer fundamental questions about the geology and geophysics of Mars \citep{Zuber:1992mola}. Its secondary objective was to measure Mars's surface reflectivity at 1064 nm to determine the varying atmospheric opacity of Mars due to dust and clouds and changes in surface albedo related to Mars's polar caps and dust storm activity \citep{Zuber:1992mola}. While the initial instrument was lost on Mars Observer in 1993, a replacement was sent to Mars on board Mars Global Surveyor (MGS), which reached Mars's orbit in 1997.   

During its prime mapping mission and part of its extended mission, MOLA simultaneously observed Mars in two modes: active sounding (usually called active radiometry) and passive radiometry \citep{Smith:2001mola,Neumann:2003clouds,Sun:2006mola}. Active sounding measured the intensity and time-of-flight of a 1064 nm laser pulse emitted by the instrument and received after reflection from Mars's surface or atmosphere \citep{Smith:2001mola,Neumann:2003clouds}. The intensity of the received signal was a function of instrument performance, surface reflectivity or aerosol backscatter coefficient, and atmospheric transmission. Therefore, if the first two types of information were known, the column opacity of the atmosphere could be derived from surface returns \citep{Neumann:2003clouds}. Passive radiometry measured 1064 nm radiation coming from the planet during the time interval between laser pulses \citep{Sun:2006mola}. This radiation originated from incoming solar radiation reflected from the surface or scattered by atmospheric aerosols. 

MOLA made 640 million topographic measurements using active sounding and an additional 176 million passive radiometry measurements. The active measurements were used to construct a global topographic map at a resolution of $1/64^{\circ}$ latitude $\times 1/32^{\circ}$ longitude \citep{Smith:2001mola}. In addition, the MOLA team published 1064 nm reflectivity maps of Mars, one using active sounding and the other using passive radiometry \citep{Smith:2001mola,Sun:2006mola}.

Both published reflectivity maps were constructed by highly approximate methods and likely contain significant biases. The active sounding reflectivity map only used data up to the end of Mars Year (MY) 24 (in the sense of \citet{Clancy:2000marsyear}) \citep{Smith:2001mola}. Due to an oversight during instrument design, the MOLA detector could not measure received power over a certain threshold. Therefore, over brighter terrains and under clear conditions, MOLA could only measure a lower bound for reflectivity. During the course of the mission, the laser pulse intensity of the instrument decreased \citep{Neumann:2003clouds}, raising the maximum detectable reflectivity. Therefore, only using data from part of the mission would limit the range of reflectivities that could be measured by active sounding. In addition, the active sounding reflectivity map did not account for attenuation of the laser pulse by atmospheric aerosols \citep{Smith:2001mola}, thereby convolving the surface reflectivity with any reduction in transmission along the laser path by aerosols. The effect of this approximation would be to underestimate reflectivity over terrains with significant aerosol opacities. \citet{Ivanov:2000PhD} proposed an algorithm to correct for aerosol opacity using collocated measurements by the Thermal Emission Spectrometer (TES) on board MGS, but this algorithm was not applied to the full dataset.

The passive radiometry reflectivity map published by \citet{Sun:2006mola} made use of most of the final dataset. However, this map was constructed using a Lambert model of reflectivity, which neglects any dependence of the reflectivity on emission angle or phase angle. These effects are typically accounted for in studies of Mars in reflected sunlight \citep{Bell:1999hst,Cantor:2007my25gds}. In addition, the effects of aerosol opacity were neglected, potentially biasing reflectivity upward or downward, depending on aerosol type and viewing geometry. 

In some cases, biases are obvious in the published maps. Active sounding reflectivity over southern Amazonis Planitia ($0^{\circ}$--$30^{\circ}$ N,$160^{\circ}$--$180^{\circ}$ W) is estimated to be  10--12\%. This area has a broadband 300--3000 nm albedo of $\sim$ 30\% \citep{Christensen:1988} and a reflectivity near 1042 nm of $>$ 30\% \citep{Bell:1999hst}. Unless there is a strong narrowband absorption at 1064 nm itself, the estimated reflectivity is biased quite low. In the case of passive radiometry reflectivity, two biases are apparent. The published reflectivity is up to 60\% brighter than observed near 1042 nm and similarly much brighter than estimated from active sounding \citep{Sun:2006mola}. While some of this bias may be due to instrumental error (passive radiometry data was not calibrated) \citep{Sun:2006mola}, the relative impacts of clouds, lack of calibration, and viewing geometry have yet to be disentangled.    

The purpose of this study is to derive a reference reflectivity map of Mars's surface at 1064 nm to create a horizontal resolution column opacity product for Mars at horizontal resolutions much higher than daily weather mapping cameras such as the Mars Orbiter Camera (MOC) on board MGS \citep{Cantor:2001ds}. Both the process of developing such a map and the resulting column opacity product are intended to support the development of future Martian orbital lidars.

\section{Data and Methods}
\label{S:2}
\subsection{Methodological Approach}
The reflectivity (used here interchangeably with albedo) of a surface at a given wavelength, $\lambda$, depends on observational geometry (the angle of incident radiation, $i$; the angle at which reflected radiation emerges, $e$; and the phase angle, $g$) and the intrinsic properties of the surface. 

This dependence can be approximated by photometric models, such as the Lambert and Minnaert models \citep{Bell:1999hst, Esposito:2006pfs}. In the Lambert model, the reflected radiance ($I$) from a surface is a function of the incident radiation ($J$), $i$, and an intrinsic property of the surface known as the Lambert albedo, $A_L$.

\begin{equation}
\label{eq:lambert}
I=JA_L\cos{i}
\end{equation}

The reflectivity, $R_{\lambda}$, is then:

\begin{equation}
\label{eq:lambertr}
R_{\lambda}=I_{\lambda}/J_{\lambda}=A_{L,\lambda}\cos{i}
\end{equation}

In the special case of $i=0$, reflectivity reduces to the Lambert albedo, a parameter that is also called "the normal albedo," hereafter denoted as $R^N$. 

In the Minnaert model, the effects of $e$ and $g$ as well as $i$ are considered:

\begin{equation}
\label{eq:minnaertr}
R^M_{\lambda}=B_\lambda(g)[\cos{i}]^{k_{\lambda}(g)}[\cos{e}]^{k_{\lambda}(g)-1}
\end{equation}

The normal albedo/reflectivity is then $B_\lambda$ at $g=0$, hereafter denoted as $R^{M,N}$.

The normal reflectivity is a desirable parameter for a reference map, because it should be the reflectivity at the approximate observational geometry of MOLA active sounding. These measurements were at nadir, so $i$, $e$, and $g$ are $\approx$ $0^{\circ}$ (typically $< 0.5^{\circ}$). The Minnaert model is specifically adopted here, because \citet{Bell:1999hst} showed that Hubble Space Telescope (HST) observations of Mars's reflectivity at 1044 nm can be modeled by a Minnaert model with a globally-fixed and phase-independent $k$ parameter of 0.7 and also characterized the phase dependence by deriving a phase parameter, $b$ such that:

\begin{equation}
\label{eq:bellcorr}
B_\lambda(0)=10^{b*g/2.5}B_\lambda(g)
\end{equation} 

Ideally, $R^{M,N}_{1064}$ (the normal Minnaert reflectivity at 1064 nm) would be derived from active sounding, the derived values would be used to identify low opacity passive radiometry observations with which to characterize the phase dependence and/or adjust the estimate of normal reflectivity, and the final $R^{M,N}_{1064}$ would be validated against the 1042 nm normal reflectivity maps of \citet{Bell:1999hst}. However, this approach is unworkable for four reasons. First, as will be shown, MOLA detector limitations and uncertainties about the reflectivity of Mars's surface and atmosphere near zero phase angle limit reliable estimates of $R^{M,N}_{1064}$ from active sounding to a small portion of the planet. Second, \citet{Bell:1999hst} suggests that characterizing the phase dependence of reflectivity is mostly unnecessary. Phase dependence is generally weak, especially for surfaces near the poles, and has an inconsistent relationship with reflectivity. The darkest and brightest surfaces brighten at low phase angle, while intermediate surfaces slightly darken. \citet{Bell:1999hst} suggests that this poor correlation between reflectivity and phase is likely due to scattering by atmospheric aerosols. Third, the passive radiometry data potentially contains a 60\% bias \citep{Sun:2006mola}. Attempting to characterize and then correct for the phase dependence of data would complicate correcting the data for this bias.      

The general approach of this study therefore was to:

\begin{enumerate}
\item Derive $R^{M,N}_{1064}$ from active sounding and opacity information from TES where possible and make a reasonable estimate elsewhere.
\item Estimate an approximate column opacity from active sounding.
\item Derive $R^{M}_{1064}$ from the passive radiometry but exclude observations made under high opacity and high phase angle conditions.
\item Assume that $R^{M}_{1064}$ is $\approx$ $R^{M,N}_{1064}$
\item Validate active and passive radiometry-based estimates of $R^{M,N}_{1064}$ against one another and HST-derived values (where possible). 
\item Re-calibrate $R^{M,N}_{1064}$ estimated from passive radiometry based on the average HST-derived $R^{M,N}$  of \citet{Bell:1999hst} (centered in a broad band near 1044 nm but equivalent to $R^{M,N}_{1064}$) over large areas of homogeneous reflectivity.
    
\end{enumerate}
  
\subsection{The MOLA Dataset}
All MOLA data was downloaded from NASA's Planetary Data System (PDS). From MY 24, $L_{s}=104^{\circ}$ -- MY 25, $L_{s}=186^{\circ}$ (with some interruption), MOLA made active sounding measurements, which measured the time-of-flight and returned power of a 1064 $\pm$ 2 nm laser pulse ("shot") emitted by the instrument at nadir toward the surface \citep{Zuber:1992mola,Smith:2001mola,Sun:2006mola}. The shots had a spot size diameter of 150 m and were spaced 300 m apart. 

The key product of active sounding was $RT$, the product of the reflectivity and the two-way transmissivity of the atmosphere, which was derived from the lidar link equation \citep{Neumann:2003clouds}: \begin{equation}
\label{eq:lidarlink}
E_r=\frac{{E_t}{t_r}{A_r}}{{\pi}z_{Mars}^2}{RT}
\end{equation}

where $E_r$ was the received pulse energy, $E_t$ was the emitted pulse energy, $t_r$ was the transmissivity of the instrument's optics, $A_r$ was the area of the receiving telescope, and $z_{Mars}$ was the distance between the instrument and the surface ("the range"). The emitted and returned energies (correcting for geometric factors) were equivalent to $I$ and $J$, so, under clear conditions, the surface reflectivity (here assumed to be $R^{M,N}_{1064}$) was equivalent to $RT$ \citep{Neumann:2003clouds}. 

Shots were attenuated by the atmosphere under opaque conditions, so the column opacity of the atmosphere ($\tau_{1064}$) would be:

\begin{equation}
\tau_{1064}=-\frac{1}{2} log\bigg(\frac{RT}{R^{M,N}_{1064}}\bigg)
\label{eq:colopac}
\end{equation}

This column opacity has been assumed to be entirely due to aerosol extinction. For Mars at 1064 nm, molecular scattering and gaseous absorption are negligible \citep{Jouglet:2007hyd}. Moreover, if aerosol scattering is primarily diffuse, very few of the photons in the laser pulse scattered by aerosol will return to the receiver. That is why the optical depth in Eq. \ref{eq:colopac} is often expressed as the integral of the extinction/attenuation coefficient in the terrestrial literature \citep{Klett:1981lidar,Vaughan:2004calipso}.

From Eq. \ref{eq:colopac}, the uncertainty in column opacity is:

\begin{equation}
\label{eq:uncert1}
\delta\tau_{1064} = -\frac{1}{2}log\bigg(\frac{1+\phi_{RT}}{1+\phi_{R^{M,N}_{1064}}}\bigg)
\end{equation}

where $\delta\tau_{1064}$ is the absolute uncertainty in the column opacity, $\phi_{RT}$ is the fractional uncertainty in RT, and $\phi_{R^{M,N}_{1064}}$ is the fractional uncertainty in the surface reflectivity.

Returns were gated into four channels based on whether sufficient power was received within a certain time after the laser pulse was emitted \citep{Neumann:2003clouds}. The fastest returns were detected by Channel 1, while the slowest returns were detected by Channel 4. Slower returns suggested a longer path length and are interpreted to be due to scattering within clouds \citep{Neumann:2003clouds}.

From Mars Year (MY) 24, $L_{s}=104^{\circ}$ -- MY 28, $L_{s} =116^{\circ}$ (with some interruptions), MOLA measured Mars's brightness due to reflected sunlight at 1064 nm with a bandwidth of 2 nm. These passive radiometry measurements were made at nadir orientation and  a resolution of 0.34 km $\times$ 3 km prior to MY 25, $L_{s}=186^{\circ}$ and at an off-nadir angle of $18^{\circ}$ and at a resolution of 0.34 km $\times$ 0.5 km subsequently. The data is reported in radiance referenced to insolation received by Mars if it had a circular orbit. 

All of the data used here is restricted to observations prior to $L_{s}=186^{\circ}$ of MY 25, because it is only during this period that active sounding and passive radiometry occur simultaneously. This data selection allows active sounding to be used to estimate atmospheric opacity during passive radiometry. For almost all of the selected period, passive radiometry is available from MOLA channel 2 only \citep{Sun:2006mola} and so only Channel 2 passive radiometry was used.

\subsection{The TES Dataset}

Absorption-only dust and water ice column opacity retrievals from nadir observations by TES on board MGS were downloaded from the PDS \citep{Smith:2004tes}. These retrievals were based on averaged radiance measurements from three pairs of two detectors that formed observational tracks spaced $\sim$ 3 km apart on the surface and displaced $\sim$ 15 km to the left of the MOLA observational track. TES nadir retrievals were not made over areas with poor thermal contrast between the surface and the atmosphere, which limited retrievals at night and over the winter pole \citep{Smith:2004tes}, so the data presented here is thus restricted to dayside (8--22 Local Solar Time (LST)) retrievals during MOLA active sounding. However, an improved nightside retrieval algorithm has been applied successfully to the TES dataset by \citet{Pankine:2013}.

The conversion factors between TES thermal infrared opacity and MOLA opacity at 1064 nm are 2.6 for dust and $\sim$ 3.0 for ice, which account for an approximate 2:1 ratio for aerosol absorption between the near-infrared and the thermal infrared \citep{Clancy2003TESepf} as well as for MOLA measuring extinction opacity as opposed to absorption opacity \citep{Montabone:2015opac}. The latter effect varies with absorption opacity, so the conversion factors are most appropriate for TES opacities less than 0.5 \citep{Smith:2004tes}. If these conversion factors are used, however, TES opacities much greater than 0.5 likely will be high enough to extinguish MOLA laser pulses \citep{Neumann:2003clouds}.

\subsection{1042 nm Reflectivity Map of Mars}

\begin{figure}[ht]
\centering\includegraphics[width=1\linewidth]{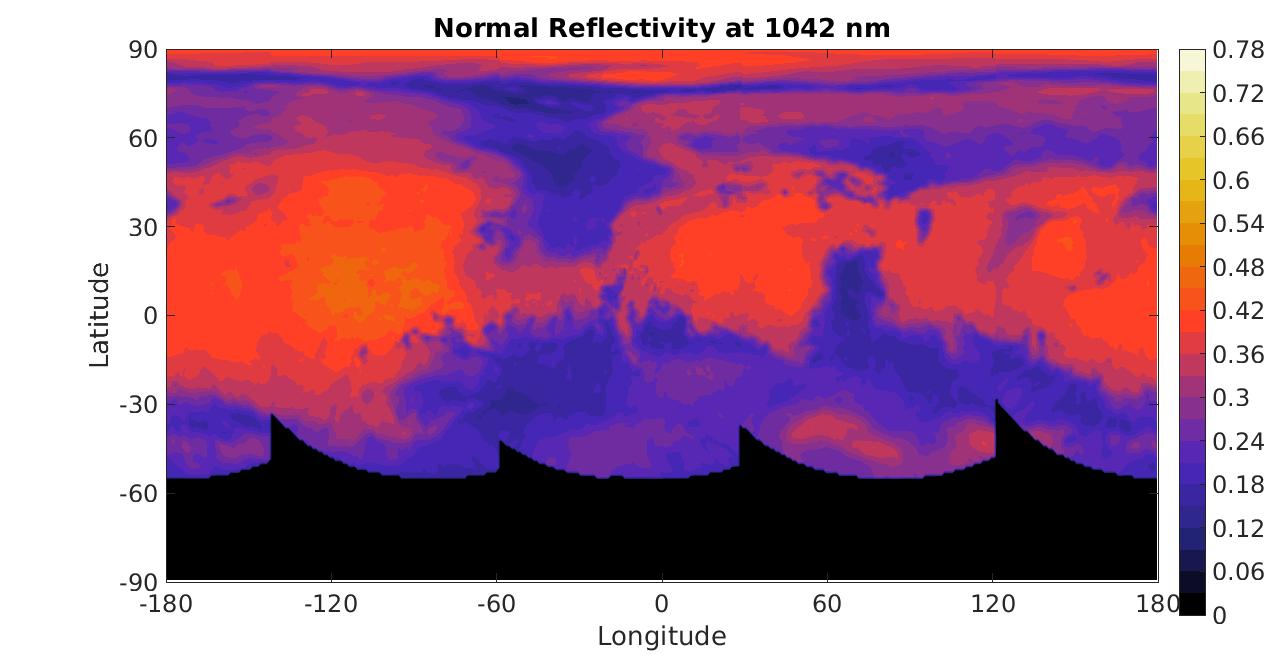}
\caption{$1^{\circ} \times 1^{\circ}$ map of estimated 1042 nm normal reflectivity from HST observations during March 1997 \citep{Bell:1999hst}. Zero reflectivity indicates areas in shadow at the time of observation.}  
\label{fig:hubblemap}
\end{figure}

A $1^{\circ} \times 1^{\circ}$ map of normal reflectivity at 1042 nm (mean $\lambda$=1018.35 nm, bandpass=36.5 nm) was obtained from J. Bell (personal communication). This map was derived from HST observations during March 1997 ($\approx L_s=90^{\circ}$, MY 23) as described by \citet{Bell:1999hst} (Fig. \ref{fig:hubblemap}). The spectrum of non-icy surfaces on Mars is featureless to $\pm$ 5\% from 600--1200 nm \citep{McCord:1982vnir,Appere:2011ice}, so significant differences in reflectivities derived from these HST observations and MOLA cannot be explained by the differences in wavelength alone. 

\subsection{Estimating Reflectivity from Active Sounding}

\subsubsection{Initial Approach}
The active sounding measurements analyzed were restricted to laser pulse returns detected by MOLA Channel 1 within 92.1 m above the surface and 81 m below the surface (as determined by the difference in the shot planetary radius and mid-point frame planetary radius in the MOLA Precision Experiment Data Record (PEDR). These bounds encompass all channel 1 returns in file number 363 within $\pm 2\sigma$ of the mean and will be used throughout this study to define a surface return. In areas of extreme slope, surface returns may be excluded by this restriction. This restriction minimizes data affected by scattering by clouds or by artifacts related to laser pulse attenuation or instrument noise \citep{Neumann:2003clouds}.

For computational convenience, the location of each measurement was rounded to its approximate location on a $0.1^{\circ} \times 0.1^{\circ}$ global grid. The reported $RT$ values were not corrected for slight variations of observational geometry during active sounding, because there is no independent way to estimate phase dependence of the Minnaert albedo. The value of $i$ and $e$ are typically $< 0.5^{\circ}$ and as much as $3^{\circ}$ for a typical file. For a Minnaert $k$ parameter of 0.7, the bias due to neglecting this correction typically will be $< 0.0015 \%$ and as much as $0.5 \%$.

Under relatively clear conditions over most surfaces, laser pulse returns exceeded the digital range of the MOLA detector \citep{Neumann:2003clouds}. In these cases, the reported $RT$ was a minimum estimate. These saturated laser pulse returns have values of raw returned power in the MOLA PEDR of 255. 

%% figure(s) showing (a) RT vs. L_s (b) 1/PW vs. L_s (c) 1/PW vs. RT (d) full link equation prediction vs. RT

\begin{figure}[ht]
\centering\includegraphics[width=1\linewidth]{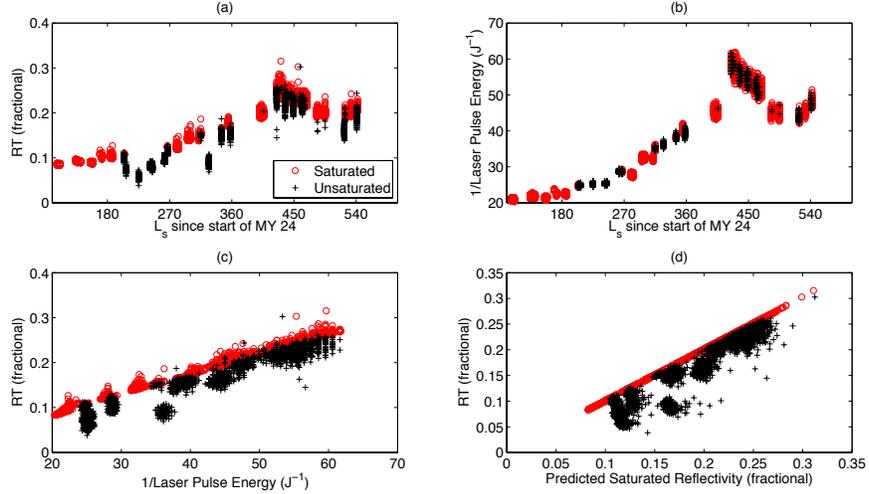}
\caption{Relationship between $RT$ and other parameters over Amazonis Planitia ($36^{\circ}$ N, $161^{\circ}$ W): (a) Variability in $RT$ over the course of MOLA active sounding; (b) Variability in the inverse laser pulse energy over the course of MOLA active sounding; (c) Relationship between inverse pulse energy and $RT$; (d) Relationship between $RT$ and an idealized saturated reflectivity derived empirically from the link equation (Eq. \ref{eq:lidarlink2}).}  
\label{fig:satdemo}
\end{figure}

The effects of saturation on $RT$ are illustrated in Fig. \ref{fig:satdemo} for a $1^{\circ} \times 1^{\circ}$ latitude-longitude bin in the bright terrain of Amazonis Planitia. Most returns are saturated. The main exceptions are contemporary with the two main periods of regional dust storm activity during MY 24 \citep{Wang:2014ds}, suggesting that the laser pulses are being attenuated by dust here. In general, though, the surface is so bright that the returned laser pulses saturate the detector (Fig. \ref{fig:satdemo}a). However, the saturated RT value varies during the course of the observations, increasing from $< 0.1$ during the summer of MY 24 to $\approx 0.25$ by the summer solstice of MY 25 and slightly decreasing thereafter(Fig. \ref{fig:satdemo}a). This secular trend is typical of bright surfaces and looks like the inverse of the daily average MOLA laser energy in Fig. 1a of \citet{Neumann:2003clouds}. The temporal evolution of inverse laser pulse energy for saturated returns strongly resembles the temporal evolution of the saturated RT returns (Fig. \ref{fig:satdemo}b). And the inverse pulse energy and RT are approximately linearly proportional for saturated returns (Fig. \ref{fig:satdemo}c). This approximately linear relationship is a consequence of Eq. \ref{eq:lidarlink}, for it is trivial to show that:

\begin{equation}
\label{eq:lidarlink2}
E_t^{-1}\frac{E_r{\pi}{z_{Mars}^2}}{{t_r}{A_r}}={RT}
\end{equation}

The terms in the quotient in Eq. \ref{eq:lidarlink2} vary in such a way to be almost constant locally. MGS orbited at roughly constant areocentric radius, so variations in $z_{Mars}$ are small for the same location on the planet. The properties of the receiving telescope did not vary significantly over the course of the mission \citep{Sun:2006mola}, so ${t_r}$ and ${A_r}$ should not vary much. All other variables being equal, $E_r$ could have changed during the mission because of in-flight changes to the threshold setting of Channel 1 \citep{Neumann:2003clouds}. Possible variations in the receiver notwithstanding, the sensitivity of MOLA active sounding measurements to variations in reflectivity and/or atmospheric opacity mainly varied throughout the mission in conjunction with laser pulse energy. 

This point can be more fully demonstrated by adapting Eq. \ref{eq:lidarlink2} to predict what the saturated value of RT should be. (Raw received power and received power are linearly proportional in most cases.) To do so, a linear regression was performed between $E_t^{-1}E_r{z_{Mars}^2}$ and $RT$ for all saturated returns meeting the data selection criteria in 4 sets of 5 individual MOLA PEDR files (of which there are 771 total). The results in Table \ref{table:satmodel} suggest that a predictive model for saturated returns is indeed possible. The slope of the model varies by only 4.3\% over the course of the mission, and the absolute value of the intercept is no more than 1.5\% of the characteristic reflectivity (0.20) \citep{Neumann:2003clouds}. A consensus value of 29 is adopted for the slope and the intercept is assumed to be 0. The consensus value of the slope should be equivalent to $\frac{\pi}{{t_r}{A_r}}$. However, the estimate of this factor from the parameters provided by the MOLA team is 32.7 \citep{Abshire:2000mola,Sun:2006mola}, which suggests that the PEDR was generated with different assumptions about receiver characteristics than reported by the team in the literature.

\begin{table}[ht]
\centering
\begin{tabular}{l l l l l}
\hline
\textbf{File Numbers} & \textbf{Slope} & \textbf{Intercept} & \textbf{$R^2$} & \textbf{$n$} \\
\hline
1--5 & 29.19 & $-1.1*10^{-4}$ & 1 & 3154567 \\
384-388 & 29.98 & $8.0*10^{-4}$ & 0.999 & 1589262 \\
600--604 & 29.00 & $0.0011$ & 1 & 2336993\\
767--771 & 28.68 & $0.0031$ & 0.999 & 668610 \\
\hline
\end{tabular}
\caption{Results of the linear regression analysis to determine a predictive model for saturated RT values.}
\label{table:satmodel}
\end{table}

This predictive model provides minimal insight for the example in Fig. \ref{fig:satdemo}d, because the unsaturated returns are lower on both the ordinate and the abscissa than the 1:1 line of saturated returns. (Bright, intermediate, and dark will be used to refer to the qualitative reflectivity of surfaces.) Here, the reflectivity cannot be determined, though the RT of the maximum saturated value ($\approx$ 0.31) is a lower bound.

\begin{figure}[ht]
\centering\includegraphics[width=1\linewidth]{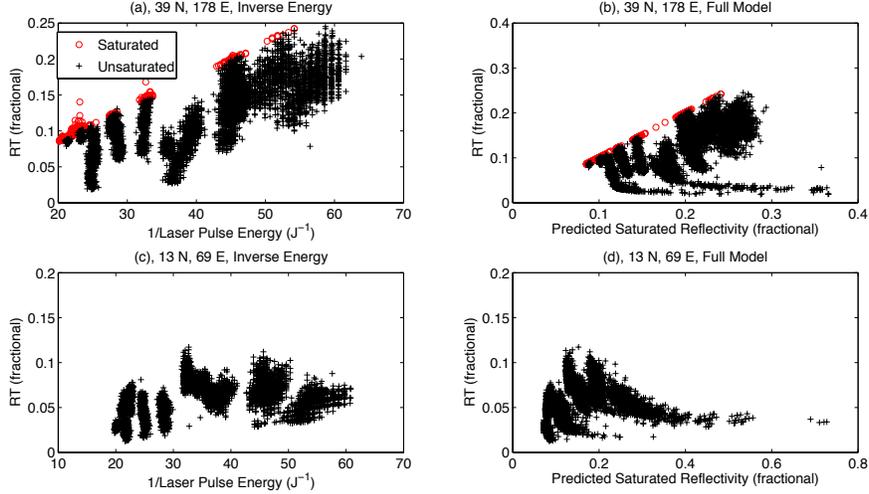}
\caption{Estimating reflectivity from active sounding over surfaces of intermediate brightness (a--b) and low brightness (c--d). The form of the plots is identical to Fig. \ref{fig:satdemo}c--d but for the $1^{\circ} \times 1^{\circ}$ bins labeled.}  
\label{fig:actrefdemo}
\end{figure}

For surfaces of intermediate brightness, however, such a plot is quite useful. On the ordinate, to the right of the saturated values, there is a region of unsaturated values, a few of which exceed the saturated values on the abscissa (Figs. \ref{fig:actrefdemo}a--b). The emergence of such a region is possible mainly because of the decrease in laser pulse energy during the course of the active radiometry measurements. While decreased laser pulse energy reduces the sensitivity of active sounding under opaque conditions, it enables the measurement of higher $RT$ values. The highest $RT$ measured in this unsaturated region is therefore a reasonable estimate for the reflectivity of this surface during a low opacity period ($\approx 0.23$ for the example in Fig. \ref{fig:actrefdemo}a--b). The minimum dayside TES opacity in this location was 0.14 (in MOLA-equivalent units). Thus, the estimated reflectivity (based on Eq. \ref{eq:colopac}) is 0.30. For surfaces so dark that all returns are unsaturated, the highest measured RT ($\approx 0.12$) is also a reasonable estimate for the surface reflectivity under low opacity conditions (Fig. \ref{fig:actrefdemo}c--d). The minimum dayside TES opacity in this location was 0.25 (in MOLA-equivalent units), so the estimated reflectivity is 0.20.  

A quick comparison with Fig. \ref{fig:hubblemap} suggests that the reflectivity estimates in the previous paragraph are somewhat higher than observed. However, positive bias is to be expected. First, $RT$ itself varies significantly, even when observing the same location at the same time (Fig. \ref{fig:satdemo}a). Part of this variability is uncertainty in the $RT$ measurement \citep{Neumann:2003clouds}. Another part of this variability could be due to reflectivity variations at the MOLA spot size scale. Second, MOLA may resolve clear areas not captured by lower resolution TES opacity data. Third, if opacity is lower at night, the minimum TES opacity will be an overestimate, because it relies on dayside data. All of these factors will result in positive bias.      

\subsubsection{A Refinement to the Initial Approach}

This positive bias was reduced by analyzing the full range of unsaturated $RT$ data available. To do so, a series of MOLA observations in the same latitude-longitude bin (e.g., $1^{\circ} \times 1^{\circ}$) and in the same 0.01$^\circ$ of $L_s$ interval were defined as an "event." A series of TES observations will occur in the same latitude bin but slightly to the west (on the dayside) and so can be treated as part of the event, because they overlap the MOLA observations in latitude and time, if not fully in longitude. It was noted that $RT \exp{(2\tau_{1064})}$ is an estimate of $R^{M,N}_{1064}$, if $RT$ is unsaturated. Using just one $RT$ value and one opacity value to characterize $R^{M,N}_{1064}$ ignores most of the available information. However, TES and MOLA data differ significantly in resolution and slightly in location. 

The differing observational patterns of MOLA and TES were addressed by generating cumulative distribution functions for opacity and $RT$ based on unaveraged data during events when MOLA and TES data are both available. In each latitude-longitude bin, this unaveraged data populated two data vectors, $\mathbf{B}(\tau)$ and $\mathbf{B}(RT)$. To better align the distributions under low opacity conditions, TES opacity data was included in $\mathbf{B}(\tau)$ in proportion to the ratio of unsaturated to saturated MOLA active sounding returns, rounding down. Thus, if there were 5 TES opacities (for example) in an event, 60 unsaturated MOLA $RT$ values, and 40 saturated MOLA $RT$ values; the 3 highest TES opacities would be added to $\mathbf{B}(\tau)$ and 60 unsaturated MOLA $RT$ values were added to $\mathbf{B}(RT)$.

From $\mathbf{B}(\tau)$ and $\mathbf{B}(RT)$ were generated cumulative distribution functions, $\mathbf{C}(\tau)$ and $\mathbf{C}(RT)$. Estimates of $R^{M,N}_{1064}$ then were drawn from $\exp[(2\mathbf{C^{-1}}(\tau)] \mathbf{\tilde{C}^{-1}}(RT)$, where $\mathbf{C^{-1}}$ refers to an inverse cumulative distribution function and $\mathbf{{\tilde{C}^{-1}}}$ refers to an inverse cumulative distribution function with a reversed order of elements. To better align the distributions under high opacity conditions, the mean of $\exp[(2\mathbf{C^{-1}}(\tau)] \mathbf{\tilde{C}^{-1}}(RT)$ was estimated and then used as a guess for $R^{M,N}_{1064}$. Then, $\min{\mathbf{B}(RT)}$ was used to estimate the TES opacities that MOLA could have not observed because of laser pulse absorption and to exclude these opacities from $\mathbf{B}(\tau)$. A guess then was made for $R^{M,N}_{1064}$ based on the new distribution of opacities. This process was iterated twice to converge on a final estimate for $R^{M,N}_{1064}$. The standard deviation of $\exp[(2\mathbf{C^{-1}}(\tau)] \mathbf{\tilde{C}^{-1}}(RT)$ was recorded as an estimate of the uncertainty in $R^{M,N}_{1064}$.

\begin{figure}[ht]
\centering\includegraphics[width=1\linewidth]{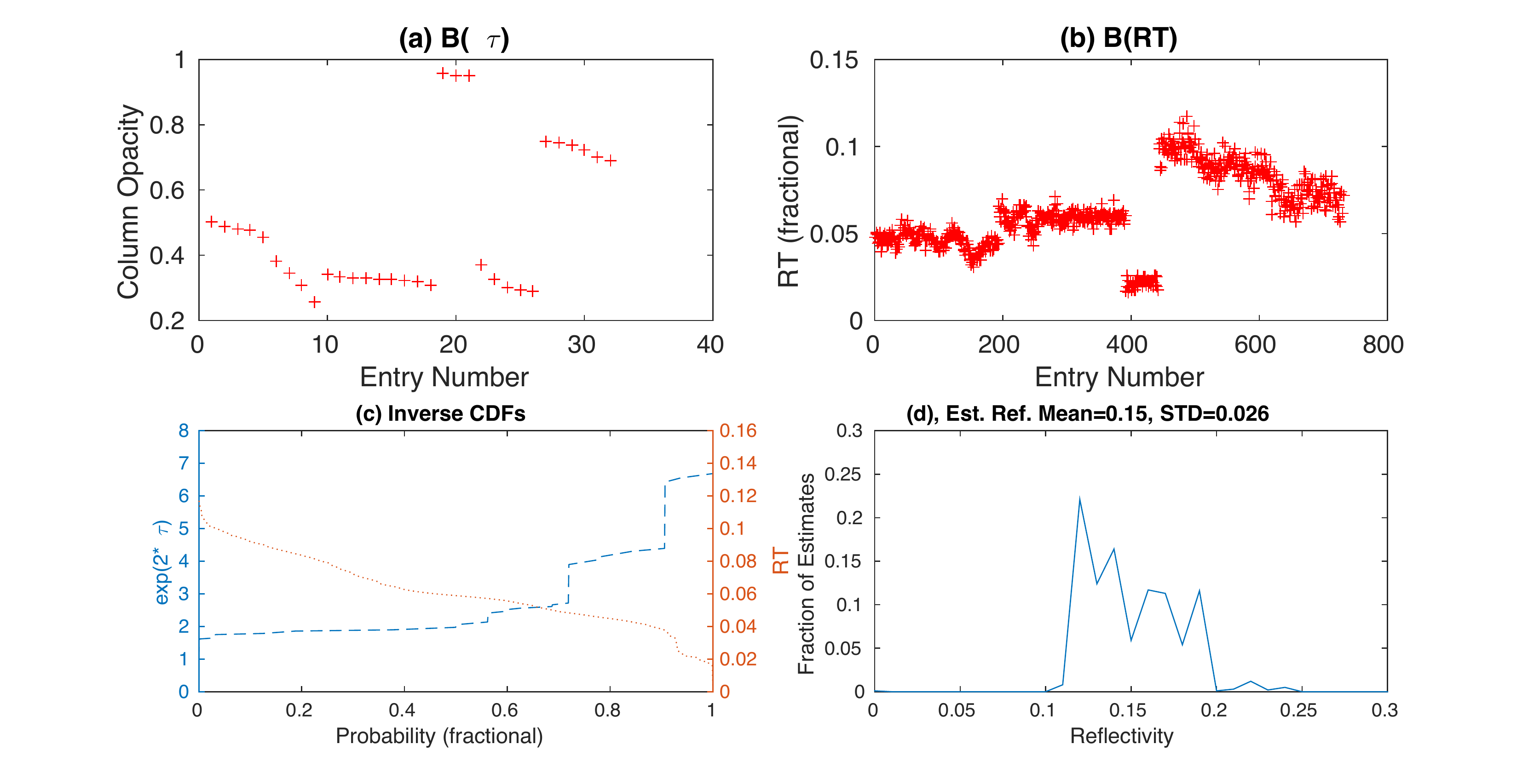}
\caption{Graphical depiction of estimating $R^{M,N}_{1064}$ over Syrtis Major ($13^{\circ}$ N, $69^{\circ}$ E) from active sounding: (a) TES opacity (in MOLA units) data vector, $\mathbf{B}(\tau)$; (b) $RT$ data vector, $\mathbf{B}(RT)$  ; (c) The inverse cumulative distribution functions, $\exp[(2\mathbf{C^{-1}}(\tau)]$ and $\mathbf{\tilde{C}^{-1}}$, whose product yields estimates of $R^{M,N}_{1064}$/. Note that multiplying values along each curve aligned parallel to the abscissa will result in a single estimate; (d) The distribution of $R^{M,N}_{1064}$ estimates based on 1000 estimates made evenly along the inverse cumulative distribution functions.}  
\label{fig:actrefdemor1}
\end{figure}

\begin{figure}[ht]
\centering\includegraphics[width=1\linewidth]{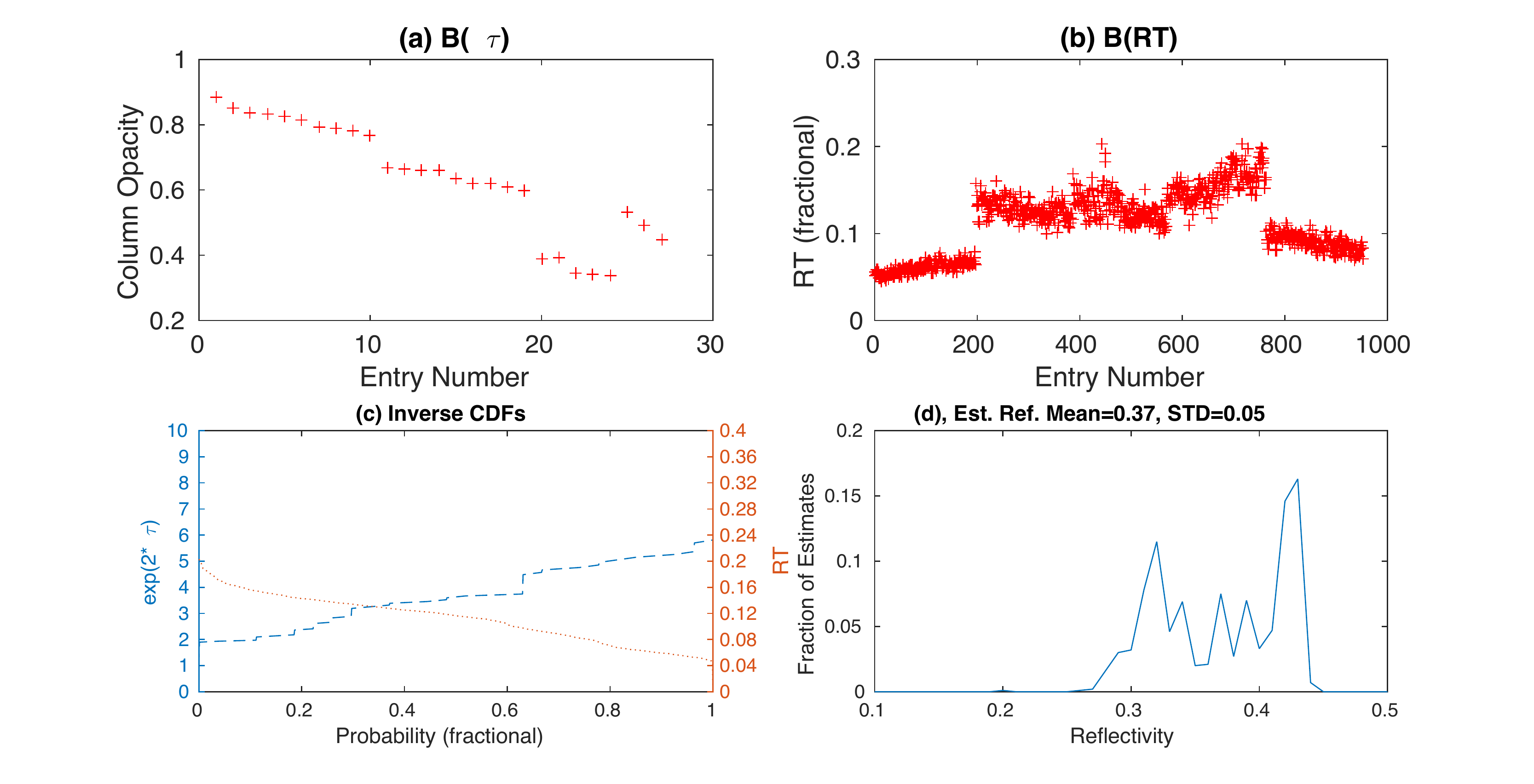}
\caption{Graphical depiction of estimating $R^{M,N}_{1064}$ over Arcadia Planitia ($38^{\circ}$ N, $178^{\circ}$ E) from active sounding. The figure is arranged as in Fig. \ref{fig:actrefdemor1}.}  
\label{fig:actrefdemor2}
\end{figure}

This algorithm is depicted graphically in Figs. \ref{fig:actrefdemor1} and \ref{fig:actrefdemor2}. These examples suggest that the example darker surface is indeed darker than the single calculation above would suggest, but the intermediate surface is brighter. Neither example suggests strong convergence of the algorithm to a single value. The main issue appears to be mismatch between the TES opacity data vector and the MOLA $RT$ data vector hinted at in Figs. \ref{fig:actrefdemor1}a--b and \ref{fig:actrefdemor2}c--d. In some cases, this mismatch is so great that the mean value $\pm$ 1.96 standard deviations is outside the physical range of reflectivity, $[0,1]$. No estimate was reported for these cases or cases in which there was insufficient data to estimate the cumulative distribution functions (5 values in each data vector).      

\subsection{Estimating Reflectivity from Passive Radiometry}
\label{S:2passrad}

Reflectivity also was estimated from passive radiometry. The passive radiometry measurements analyzed were restricted to those with solar zenith angles less than $80^{\circ}$, anomaly flags of 0, and off-nadir angles less than $1^{\circ}$. The restriction in solar zenith angle minimizes contributions from scattering near the terminator. For computational convenience, the location of each measurement was rounded to its approximate location on a $0.1^{\circ} \times 0.1^{\circ}$ global grid. All measurements were then converted to Minnaert reflectivity according to Eq. \ref{eq:minnaertr} and using a phase-independent $k$ of 0.7 \citep{Bell:1999hst,Cantor:2007my25gds}. $R^{M,N}_{1064}$ was then estimated from the statistical distribution of a selection of data (hereafter called "the select data") in a small latitude-longitude bin ($0.5^{\circ} \times 0.5^{\circ}$ or $1^{\circ} \times 1^{\circ}$). This data selection has two elements. 

%%figure
\begin{figure}[ht]
\centering\includegraphics[width=1\linewidth]{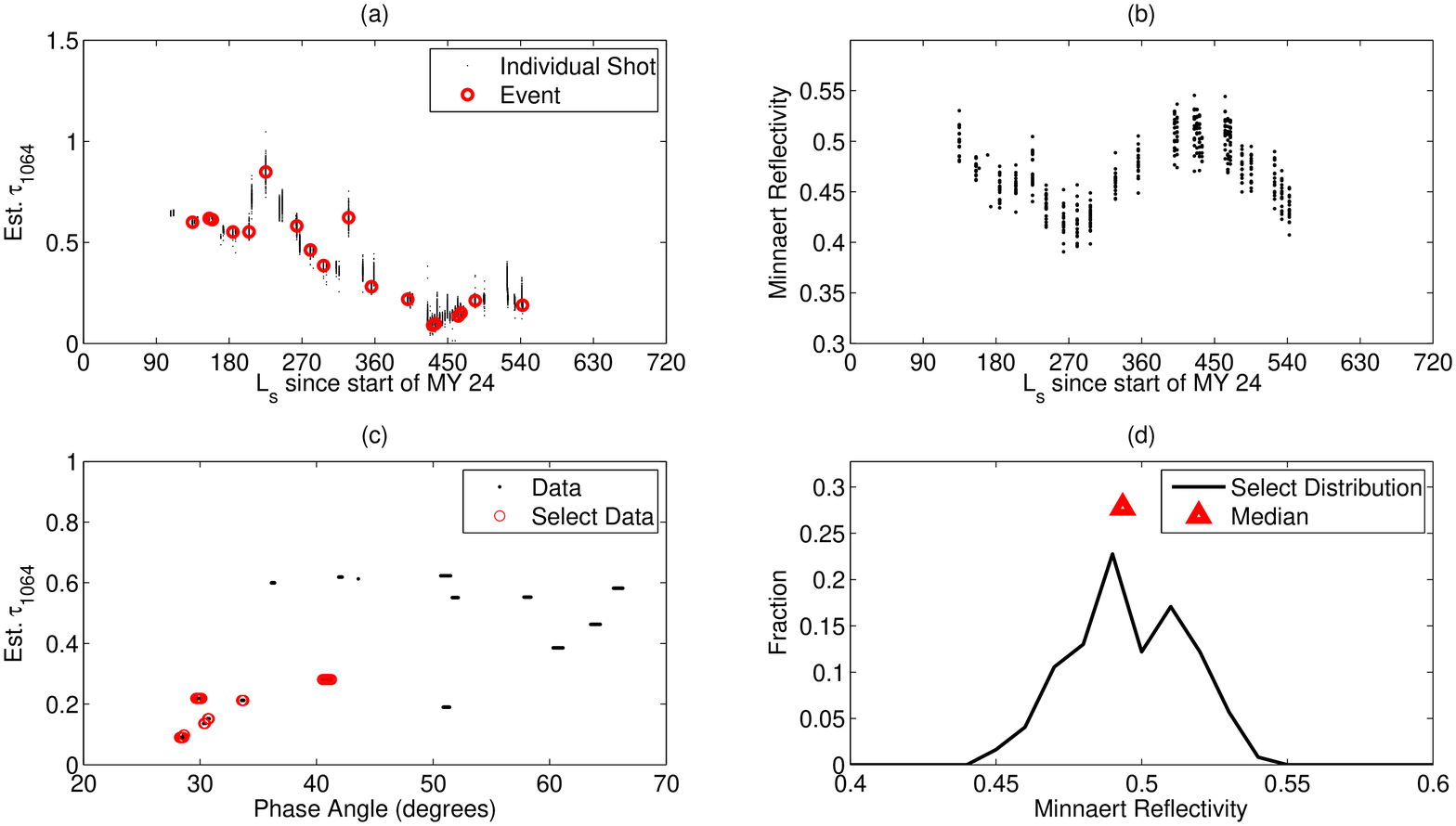}
\caption{Graphical depiction of estimating $R^{M,N}_{1064}$ over Amazonis Planitia ($36^{\circ}$ N, $159^{\circ}$ W) from passive radiometry: (a) Variation in estimated $\tau_{1064}$ from active radiometry. Columns of black dots without red circles indicate instances when appropriate passive radiometry and active radiometry do not overlap (during the night, for example); (b) Variation in Minnaert reflectivity with time; (c) The select low phase angle and low $\tau_{1064}$ data; (d) The median of the distribution of select passive radiometry measurements provides the estimate of $R^{M,N}_{1064}$. Binning is at 0.01 reflectivity unit intervals.}  
\label{fig:passraddemo1}
\end{figure}

%%figure
\begin{figure}[ht]
\centering\includegraphics[width=1\linewidth]{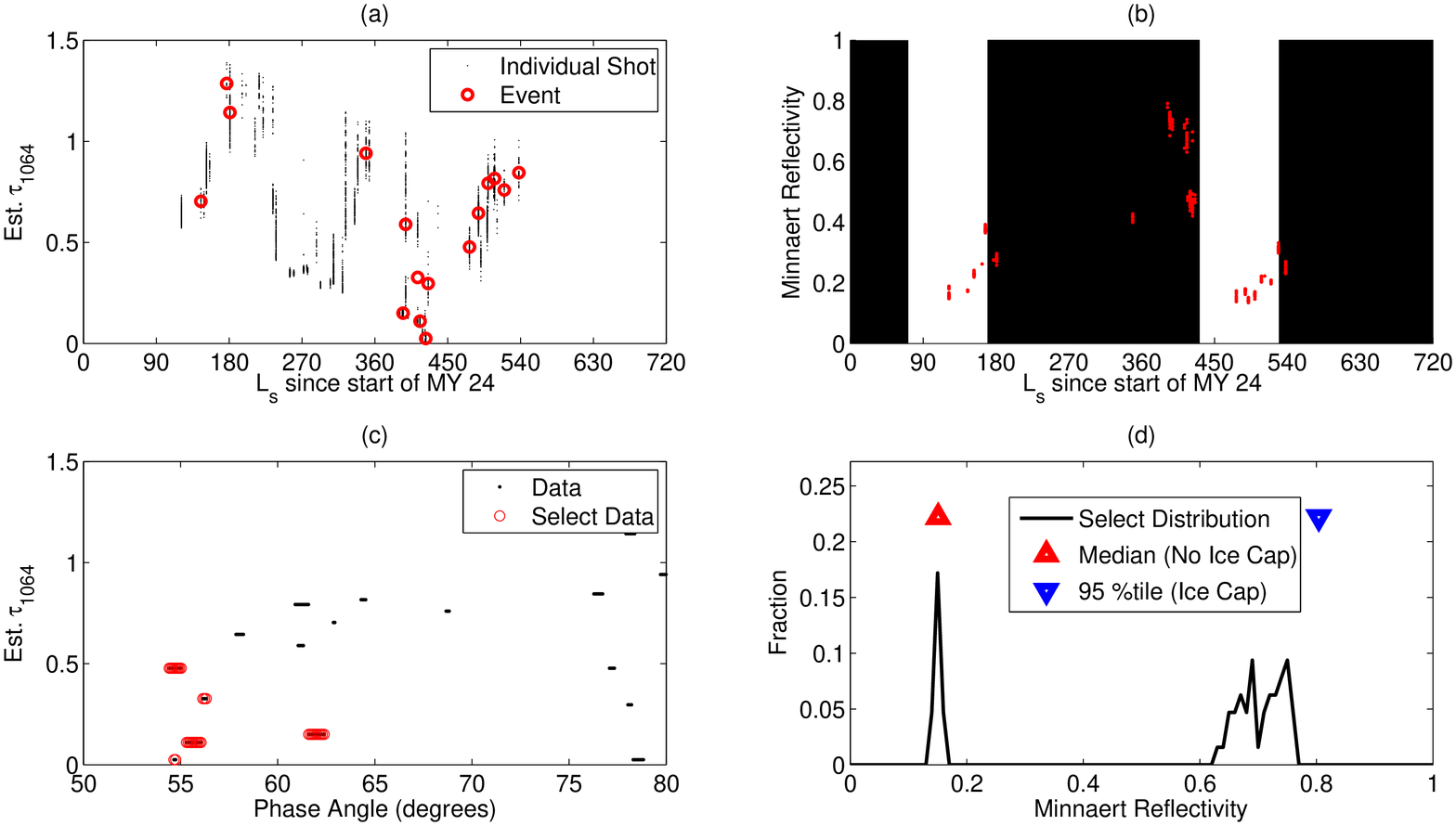}
\caption{Graphical depiction of estimating $R^{M,N}_{1064}$ near the North Pole ($76^{\circ}$ N, $0^{\circ}$ E) from passive radiometry: (a) Variation in estimated $\tau_{1064}$ from active radiometry; (b) Variation in Minnaert reflectivity with time. The black areas indicate times when the climatological seasonal $CO_2$ ice cap extended to an equivalent latitude of $76^{\circ}$ N with an uncertainty of $\pm$ 15 $^{\circ}$ of $L_s$ \citep{Piqueux:2015cap}; (c) The select low phase angle and low $\tau_{1064}$ data; (d) The median of the distribution of select passive radiometry measurements when the ice cap is absent provides the estimate of $R^{M,N}_{1064}$ while the 95th percentile value of the distribution when the cap is present is used to estimate the cap reflectivity} 
\label{fig:passraddemo2}
\end{figure}

First, for latitudes $\leq 45^{\circ}$, only measurements at $g\leq 45^{\circ}$ were selected. For latitudes $> 45^{\circ}$, only measurements at $g \leq$ the absolute value of the latitude were selected. This selection is a tradeoff. On one hand, measurements at low phase angles are preferable. The Minnaert photometric model has multiple solutions at $g > 30^{\circ}$ \citep{Esposito:2006pfs}. Moreover, the minimal $g$ dependence found by \citet{Bell:1999hst} is only known to be valid for $g < 45^{\circ}$. And since the passive radiometry measurements observe at $e \approx 0$, using low phase angle data minimizes the effects due to breakdown of the Minnaert model at high $e$ \citep{Bell:1999hst,Esposito:2006pfs}. On the other hand, low phase angle observations are rare or non-existent at high latitudes, so the phase angle filter needs to be relaxed at higher latitudes.         

Second, the data was further restricted to passive radiometry measurements when the mean $\tau_{1064}$ inferred from active sounding was $< 1.0$ (Fig. \ref{fig:passraddemo1}a). The mean $\tau_{1064}$ is evaluated over all active sounding measurements in each intersection of the MOLA observational footprint with the small area used for binning (an observational event, in the sense used in the previous sub-section). For binning at $0.5^{\circ} \times 0.5^{\circ}$, $\approx$ 5--20 events take place near the Equator and $\approx$ 100 at 80$^{\circ}$ N and S. Each event contains $\approx$ 10 measurements at $0.5^{\circ} \times 0.5^{\circ}$ binning. Where $R^{M,N}_{1064}$ can be estimated from active sounding, this estimate is used to calculate $\tau_{1064}$ based on Eq. \ref{eq:colopac}. In all other cases, $R^{M,N}_{1064}$ is assumed to be 0.46 (the upper bound I/F value implied by the distributions in Fig. 1  of \citet{Bell:1999hst}) in order to calculate $\tau_{1064}$.

The selection criterion for opacity takes account of present understanding of the relative contributions of atmospheric scattering and surface reflection to the apparent brightness of the surface. Figs. 7 and 8 of \citet{Clancy2003TESepf} suggest that dust and water ice only contribute $\sim$ 5\% to the apparent albedo of the surface observed at zero emission angle at 1064 nm for a relatively dark surface with visible opacity of 0.25. Figs. 6--8 of \citet{Clancy2003TESepf} taken together strongly suggest that the relative contribution of atmospheric scattering and surface reflection to TES broadband visible/near-infrared albedo and MOLA passive radiometry should be similar. 

A more detailed analysis of the effect of opacity on TES broadband albedo by \citet{Szwast:2006} established three principles: (1) the major effect of atmospheric opacity on the apparent brightness of the surface is during large-scale dust storm activity, when TES dust opacities are greater than 0.4; (2) the aphelion equatorial cloud belt has an effect on TES broadband albedo of $\sim$ 0.01 absolute; and (3) the effect of scattering by dust and water ice is important for darker than brighter surfaces. Tharsis, for example, did not brighten during the MY 25 global dust storm, while darker surfaces brightened by 0.05 absolute or more. Therefore, it is only necessary to exclude observations made under MOLA-equivalent opacities of $\approx$ 1.04 over dark and intermediate surfaces.

Analysis and modeling of the Viking Infrared Thermal Mapper (IRTM) Emission Phase Function (EPF) sequences by \citet{ClancyandLee:1991} suggest that the relative contribution of atmospheric scattering often will increase at high phase angle. The critical phase angle beyond which the atmospheric contribution rapidly increases is a function of opacity, the single scattering albedo of the aerosol, and the brightness of the surface. However, broadly speaking, sampling $\tau_{1064} < 1.04$ and $g < 45^{\circ}$ in observations near 14:00 LST will avoid signficant contributions from atmospheric scattering over non-icy surfaces. Over icy surfaces, water ice clouds tend to have a minimal impact on the apparent reflectivity for $g$ up to $90^{\circ}$ \citep{ClancyandLee:1991}. However, dust with $\tau_{1064} = 1.04$  will reduce the apparent reflectivity of icy surfaces by up to 20 \% at $g < 90^{\circ}$ \citep{ClancyandLee:1991}. Nevertheless, significant dust opacities over the polar caps are mostly observed during large-scale dust storm activity \citep{Szwast:2006}, whose contribution should be mostly excluded by opacity filtering.  

$R^{M,N}_{1064}$ is then estimated to be the median of the select data. The median is preferred to the mean because the phase angle--opacity selection process is imperfect. However, this approach also reduces error due to seasonal and/or episodic variability in surface reflectivity. 

Seasonal variability in reflectivity is a minor issue over Amazonis Planitia. Opacity varies with regional dust storm activity, as discussed earlier (Fig. \ref{fig:passraddemo1}a). Apparent reflectivity is lowest at southern summer solstice and highest at northern summer solstice. This variability is probably due to geometry rather than actual variability in reflectivity. Note that the low phase angle and low opacity data is within 10\% of the median of 0.49 (Fig. \ref{fig:passraddemo1}d). Moreover, Fig. 33 of \citet{Szwast:2006} shows TES broadband albedo in this area varied in the opposite sense in MY 24 and 25. Therefore, this variability may be due to the area appearing darker when observed at higher phase angle in winter.

Note also that the seasonal variability signal is comparable to the variance of Minnaert reflectivity during an individual event (Fig. \ref{fig:passraddemo1}b). The geometry during an individual event is equivalent to $\approx 1^{\circ}$. Therefore, either the variability in the surface reflectivity in the area or the precision of the measurements (or the combination thereof) is comparable in amplitude to seasonal variability. Note that \citet{Sun:2006mola} estimates relative errors in individual radiance measurements $<5\%$, which appears roughly consistent with the range of variability observed in a typical event (10--20\%). 

Near the poles, however, the amplitude of seasonal variability is much larger. In Figs. \ref{fig:passraddemo2}a--b, the estimated opacity and Minnaert reflectivity records of an area near the North Pole are shown, including the break in overlapping active radiometry and applicable passive radiometry measurements in polar night. Once the area emerges from polar night in MY 25, the reflectivity has increased slightly, but under cloudy conditions. When opacity decreases during northern spring, the Minnaert reflectivity from passive radiometry increases to $\approx$ 0.7 and the decreases again in cloudy conditions by $ L_s \approx 70^{\circ}$, finally decreasing to the same level as the previous northern summer (a moderately cloudy period as well).

One possibility for the rise in reflectivity during northern spring is that the selection of phase angle was too generous, resulting in a contribution from limb brightening effects as the area came once again under observation. The select data does vary somewhat in phase angle (Fig. \ref{fig:passraddemo2}c), but it is hard to imagine the difference in phase angle explaining an 0.5 unit range in reflectivity (Fig. \ref{fig:passraddemo2}d). In addition, the rise does not seem associated with atmospheric aerosol. The area appears to be clearest when it is brightest (Fig. \ref{fig:passraddemo2}a). Instead, the rise is probably due to observation of the seasonal CO$_2$ cap. This conclusion is consistent with \citet{Piqueux:2015cap}, which shows the northern seasonal cap is in place here until ${\approx} L_s=70^{\circ}$. 

However, this interpretation is not straightforward for two reasons. First, clouds are probably obscuring the cap and lowering the apparent reflectivity in passive radiometry. Second, brightening of the surface could produce an apparent clearing effect in Fig. \ref{fig:passraddemo2}a even if cloudiness did not change. For the factor of 4 increase in reflectivity implied by Fig. \ref{fig:passraddemo2}d, Eq. \ref{eq:colopac} suggests that the inferred optical depth would decrease by 0.7, which would explain much of the clearing during northern spring of MY 25. Indeed, the negative opacity in northern spring is consistent with the presence of the cap. The reflectivity inferred from active sounding here was 0.14 $\pm$ 0.02, which is close to the reflectivity when the cap is not present. (TES opacities would not have been retrieved here when the cap was present, because the surface would have been too cold.) The observation of $RT$ values much higher than this value implies the presence of a much more reflective surface. 

As a first order approximation to disentangling these effects, $R^{M,N}_{1064}$ was estimated differently in areas affected by the seasonal caps. The select data was filtered on the basis of whether the climatological seasonal cap was present at the equivalent latitude of the bin, as indicated by the analysis of \citet{Piqueux:2015cap}. To account for interannual and longitudinal variability, the cap was treated as being present for 15 $^{\circ}$ of $L_s$ before and after the climatological mean period the cap lasted. $R^{M,N}_{1064}$ is then the median of the filtered select distribution when the cap is not present. 

The 95th percentile of the filtered select distribution when the cap is present was retained as an estimate of $R^{M,N}_{1064}$ for the seasonal cap during polar night. Choosing an extremum of the distribution is justified on two grounds. First, the clouds that obscure the cap are probably less reflective than the cap itself. Second, the cap is mainly observed by passive radiometry when it is sunlit and receding, and thus when the cap may be less reflective in passive radiometry because of age or exposure of underlying surface material as the cap recedes.

\section{Reflectivity Analysis}
\label{S:3}

\subsection{Results}

Active sounding is insufficient to derive $R^{M,N}_{1064}$ globally. As suggested by Figs. \ref{fig:satdemo}--\ref{fig:actrefdemo}, the upper detection limit for RT due to the saturation of the detector is rarely greater than 0.3 and was often much less. Thus, evaluating the reflectivity of surfaces brighter than this will be highly dependent on the opacity conditions during the season when laser pulse energy was lowest (around northern summer solstice of MY 25), when the atmosphere was relatively clear away from the equatorial cloud belt.

\begin{figure}[ht]
\centering\includegraphics[width=1\linewidth]{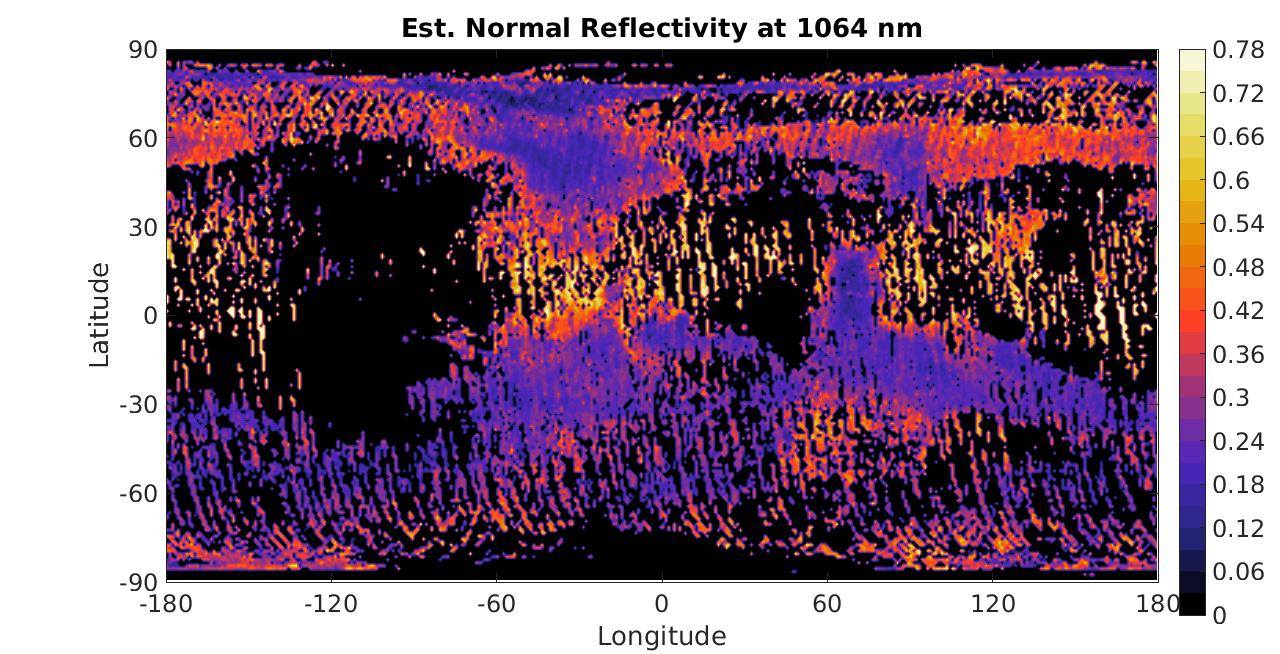}
\caption{$1^{\circ} \times 1^{\circ}$ map of $R^{M,N}_{1064}$ estimated by active sounding. Zero reflectivity indicates where no estimate is possible. The map is saturated at 0.78 to enable better comparison with Fig. \ref{fig:hubblemap}.}  
\label{fig:activemap}
\end{figure}

Thus, almost all estimates of $R^{M,N}_{1064}$ from active sounding are for darker terrains  such as Syrtis Major ($10^{\circ}$ N, $70^{\circ}$ E), Mare Acidalium ($45^{\circ}$ N, $30^{\circ}$ E), and Utopia Planitia ($45^{\circ}$ N, $90^{\circ}$ E) (Fig. \ref{fig:activemap}). The map traces out the dark terrain on the southern margin of the northern ice cap at $75^{\circ}$ N, faintly traces Gemina and Hyperborea Lingulae ($80^{\circ}$ N, 0--$60^{\circ}$ W), and reproduces the strong hemispheric dichotomy in reflectivity between the northern and southern hemispheres present in Fig. \ref{fig:hubblemap}. Darker terrains are mapped in the southern tropics and mid-latitudes, while reflectivity only can be estimated fortuitously on bright terrains of the northern tropics and mid-latitudes. 

\begin{figure}[ht]
\centering\includegraphics[width=1\linewidth]{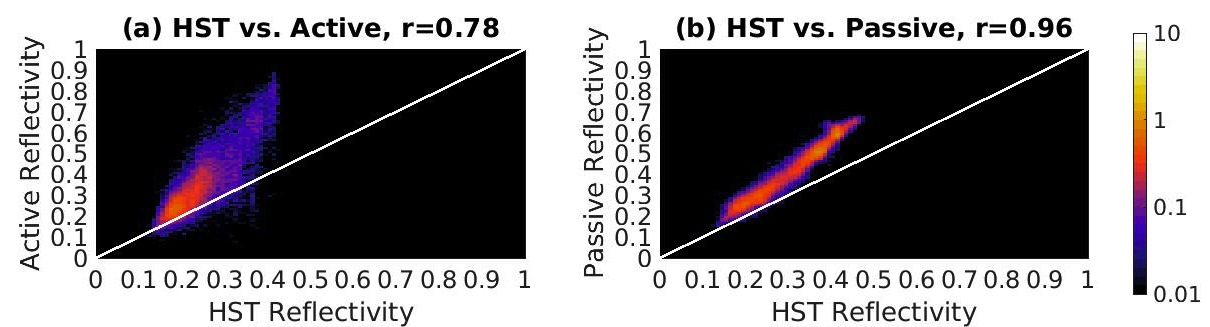}
\caption{Bivariate histograms (\%) of $R^{M,N}$ estimated from HST observations, active sounding, and passive radiometry, as labeled. These plots can be understood as x-y scatterplots with the data being compressed by binning by 0.01 units of reflectivity. The correlation coefficient between the datasets is listed in the panel titles. A white 1:1 line is in each panel. Comparison between each dataset is only made for latitude-longitude bins where the reported data is present for each dataset.}  
\label{fig:2dhists}
\end{figure}

\begin{figure}[ht]
\centering\includegraphics[width=1\linewidth]{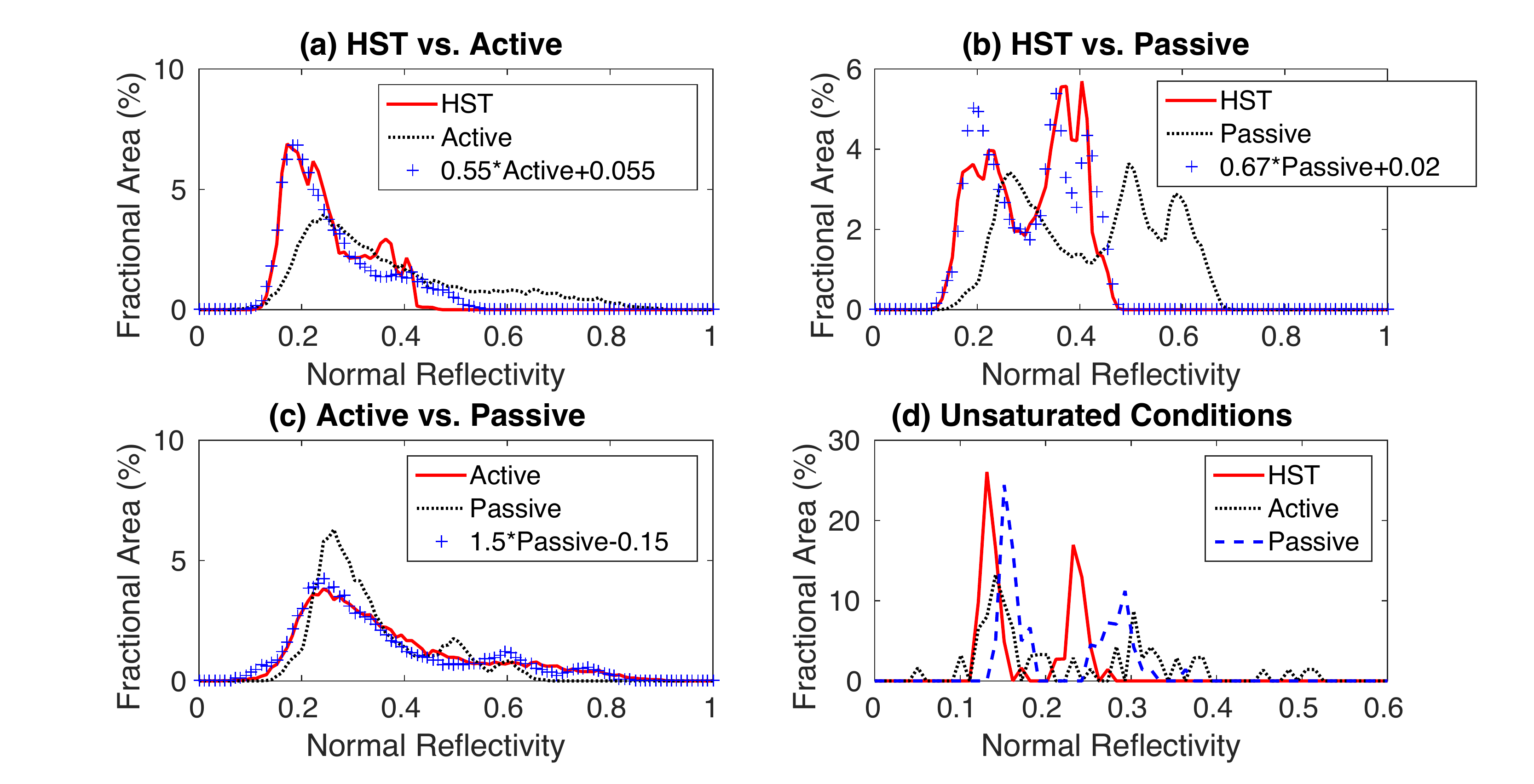}
\caption{Area-weighted distributions of $R^{M,N}$ estimated from HST observations, active sounding, and passive radiometry, as labeled. The blue crosses in a--c show what would happen if the labeled dataset were tuned by inspection to match the distribution of other dataset in the panel. Panel d shows the distributions for the latitude-longitude bins (at $1^{\circ} \times 1^{\circ}$ resolution) where all RT data was unsaturated. Comparison between each dataset is only made for latitude-longitude bins where the reported data is present for each dataset.} 
\label{fig:refdists}
\end{figure}

But despite limited data over bright surfaces, $R^{M,N}_{1064}$ derived from active sounding strongly correlates with that inferred from HST observations (Fig. \ref{fig:2dhists}a). However, active sounding overestimates $R^{M,N}_{1064}$ over all but the darkest terrains (Figs. \ref{fig:2dhists}a and \ref{fig:refdists}a). Part of this bias could be attributable to insufficient unsaturated $RT$ data. However, comparisons in areas where all data is unsaturated (the dark center of Syrtis Major and the brighter, but cloudy north rim of Hellas) suggests that the positive bias over bright terrains is real (Fig. \ref{fig:refdists}d). This bias also is present if the active sounding reflectivity analysis is performed over a bin 25$\times$ larger than the standard latitude-longitude bin in a uniformly bright region. Surfaces appear much brighter in active sounding data than in HST observations.

\begin{figure}[ht]
\centering\includegraphics[width=1\linewidth]{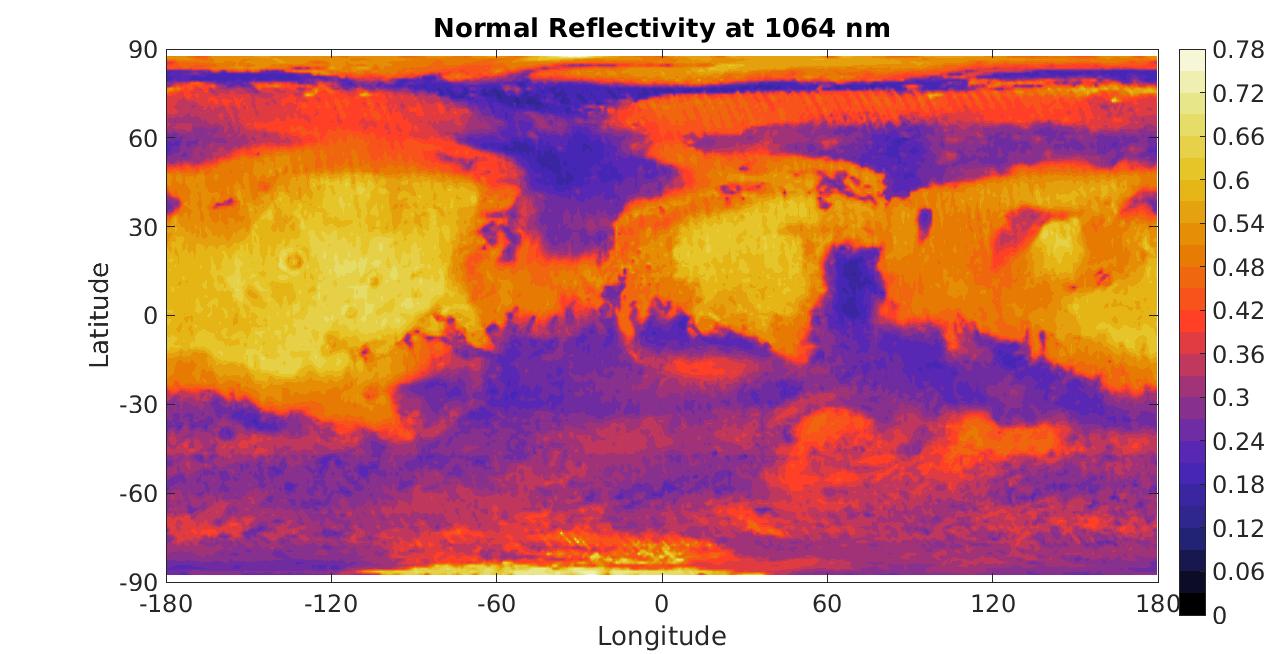}
\caption{$1^{\circ} \times 1^{\circ}$ map of $R^{M,N}_{1064}$ estimated from the median of the select passive radiometry data distribution over areas where the seasonal polar cap was absent. Zero reflectivity (black dotted texture) indicates where no estimate is possible. Any missing data was filled by interpolation.}  
\label{fig:passivemap}
\end{figure}

Passive radiometry can be used to estimate reflectivity over almost all of the planet, including the brightest areas. The example map (Fig. \ref{fig:passivemap}a) captures key features of the HST map (Fig. \ref{fig:hubblemap} and the composite map in \citet{Bell:1999hst}, including the north-south dichotomy in reflectivity, major dark terrains such as Syrtis Major, major bright terrains such as Arabia ($15^{\circ}$ N, $40^{\circ}$ E), and even minor dark albedo features such as one at ($40^{\circ}$ N, $160^{\circ}$ W). There is still some missing data, especially near the poles, which was filled by interpolation.

The map has some stripy texture, particularly at $75^{\circ}$ N across most longitudes. Having excluded the possibility of long-term drift in the reflectivity measurements, \citet{Sun:2006mola} attributed this type of texture in the reflectivity map presented there to the MGS orbit pattern: adjacent areas may have been sampled in different years, so temporal variability in reflectivity due to dust deposition or removal was aliased as spatial variability. This effect seems possible here, though the effect is easier to see at higher latitude, where variations in phase angle sampling could create a similar effect.

Another difference between the example passive radiometry map and HST observations is the magnitude of reflectivity. For example, Arabia has $R^{M,N}_{1064}$ of at least 0.6 in the passive radiometry map (Fig. \ref{fig:passivemap}), while its reflectivity is $\approx$ 0.4 in the HST data (Fig. \ref{fig:hubblemap}) \citep{Bell:1999hst}. However, over Syrtis Major, $R^{M,N}_{1064}$ is $\approx$ 0.15, which is quite similar to the HST reflectivity data.       

The correlation between $R^{M,N}_{1064}$ derived from passive radiometry and the HST-based estimate is nearly perfect (Fig \ref{fig:2dhists}b). For reflectivities $<$ 0.25, these estimates appear to differ by a constant offset of $\sim$ 0.02. This offset grows over brighter terrains to $\sim$ 0.2 (Fig \ref{fig:2dhists}b). Further evidence of the bias can be seen when comparing area-weighted distributions. Terrains would look similar in passive radiometry and HST if the former were reduced by 33 \% (Fig. \ref{fig:refdists}b). This is consistent with the 50 \% positive bias of passive radiometry-based reflectivity reported by \citet{Sun:2006mola}. Thus, surfaces generally appear brighter in passive radiometry data than in HST observations. 

\subsection{Interpretation and Re-calibration}
Surfaces appear brighter in MOLA active sounding and passive radiometry than HST observations. The positive bias of passive radiometry is unsurprising. It was reported by \citet{Sun:2006mola}. However, the bias is smaller or zero over the darkest terrains and cannot be easily explained away as a result of clouds or variations in observational geometry. These factors play a role (Fig. \ref{fig:passraddemo1}), but the bias remains when they are excluded.   

The positive bias (perhaps up to 100\%) in active sounding is more puzzling. (Note that this positive bias would result in excluding passive radiometry observations under lower opacity conditions than required.) It is unlikely to be a calibration issue. Unlike the component setup involved in passive radiometry, the MOLA receiver used for estimating the energy of received pulses in active sounding was calibrated prior to flight and did not experience significant performance issues prior to the end of active sounding \citep{Abshire:2000mola,Sun:2006mola}.

It is possible that the ratio of MOLA to TES opacities has been overestimated. At the MOLA-equivalent opacities of $\sim$ 1 under which the reflectivity of bright surfaces are generally assessed, an overestimate of 35 \% would explain a bias of 100 \%. However, such a bias would be true for all surfaces. Yet the maximum opacity in Fig. \ref{fig:actrefdemor1}a corresponds to $RT$ of $\sim$ 0.02 (Fig. \ref{fig:actrefdemor1}b), which implies a reflectivity of 0.13, an estimate at the low end of the range.

Another possible explanation is that Mars looks brighter in active sounding data than it does from a space-based telescope. An instrument like HST observes Mars at a variety of solar phase angles, while MOLA typically observed Mars's surface at 0.2$^{\circ}$--0.4$^{\circ}$ relative to the illumination of its laser beam. Therefore, MOLA observations could include two effects that are difficult to observe in any other way. 

First, MOLA could observe enhanced brightness of the Martian surface due to the Shadow Hiding Opposition Effect (SHOE) or Coherent Backscatter Opposition Effect (CBOE) \citep[e.g.,][]{Hapke:1998}. The possibility of opposition surge effects enhancing MOLA returns was mentioned by \citet{IvanovMuhleman:1998} and \citet{Smith:2001mola}. However, there remains no way to estimate the size of this effect. Observations of Mars at $< 3^{\circ}$ are rare and those under clear conditions are practically non-existent \citep{Thorpe:1978,Soderblom:2006,Vincendon:2013,Fernando:2013}. And there is not enough MOLA data at non-nominal phase angles to characterize the opposition effect either.

Second, MOLA could observe the forward scattering peak of atmospheric aerosol. Typical nadir observations of Mars aerosol are made with the Sun illuminating the aerosol from above. The sunlight is strongly forward scattered, resulting in little additional brightness being viewed at low opacity because the backscattered light is quite diffuse (scatters over a wide variety of phase angles)  \citep{Clancy2003TESepf,Wolff:2009,Dabrowska:2015}. The portion of the MOLA beam reflected from the surface, however, creates the opposite viewing geometry. This reflecting beam illuminates aerosol from below, so that the MOLA receiver observes forward scattered radiation. However, the scattering phase function in the forward scattering direction is quite direct (it has a strong peak near $g=180^{\circ}$, near where the MOLA receiver would observe it \citep{Wolff:2009,Dabrowska:2015}. 

The effect of forward scattering would be magnified when diagnosing reflectivity from active sounding over bright surfaces in two ways. First, the effect is proportional to the magnitude of the reflection from the surface (higher over bright surfaces). Second, the saturation effect forces reflectivity to be inferred from active sounding in high opacity conditions. The conversion factor of 2.6--3.0 between TES and MOLA opacity assumes diffuse scattering. Additional direct scattering by aerosol, which could be greater under high opacity conditions, would reduce the apparent extinction of the MOLA pulses. Thus, the true conversion factor would be lower. The exponential dependence of the apparent reflectivity on opacity enhances the bias. To complicate matters further, the vertical distribution and characteristics of aerosol likely would change this scattering and its contribution to a gated MOLA surface return. Such an effect would be easy to see if there were unsaturated MOLA active sounding data under a wide range of opacity conditions over the brightest surfaces. But there is not.

Nevertheless, the central purpose of this study is to derive aerosol column opacity from MOLA observations. Despite the issues outlined above, such a product is possible, though a few caveats are in order. The reflectivity derived from passive radiometry has a positive bias, but it is a monotonic one that strongly correlates with the validating dataset (HST). It is therefore possible to re-calibrate this data with the help of the HST dataset. With the Martian opposition surge near zero phase angle unknown, the estimate of $R^{M,N}_{1064}$ from passive radiometry should be acceptable to remove the effect of reflectivity from surface returns in MOLA active sounding measurements. The resulting reflectivity map is only weakly affected by assumptions about the ratio of MOLA to TES opacity. 

However, any column opacity derived from these measurements may have significant uncertainty due to atmospheric scattering. Eq. \ref{eq:uncert1} implies that the 100 \% bias in $RT$ inferred for some surfaces (Fig. \ref{fig:2dhists}a) would result in MOLA column opacities underestimating a 1064 nm opacity based on diffuse scattering by 0.35. The unphysical values excluded from the active radiometry analysis suggest larger errors could be possible. Note that opposition effects at low phase angle would impact the error budget of column opacity in the opposite direction. A 100 \% bias in apparent surface reflectivity would imply surface reflectivity was 50 \% of its true value and result in column opacity being underestimated by 0.35.

\begin{figure}[ht]
\centering\includegraphics[width=1\linewidth]{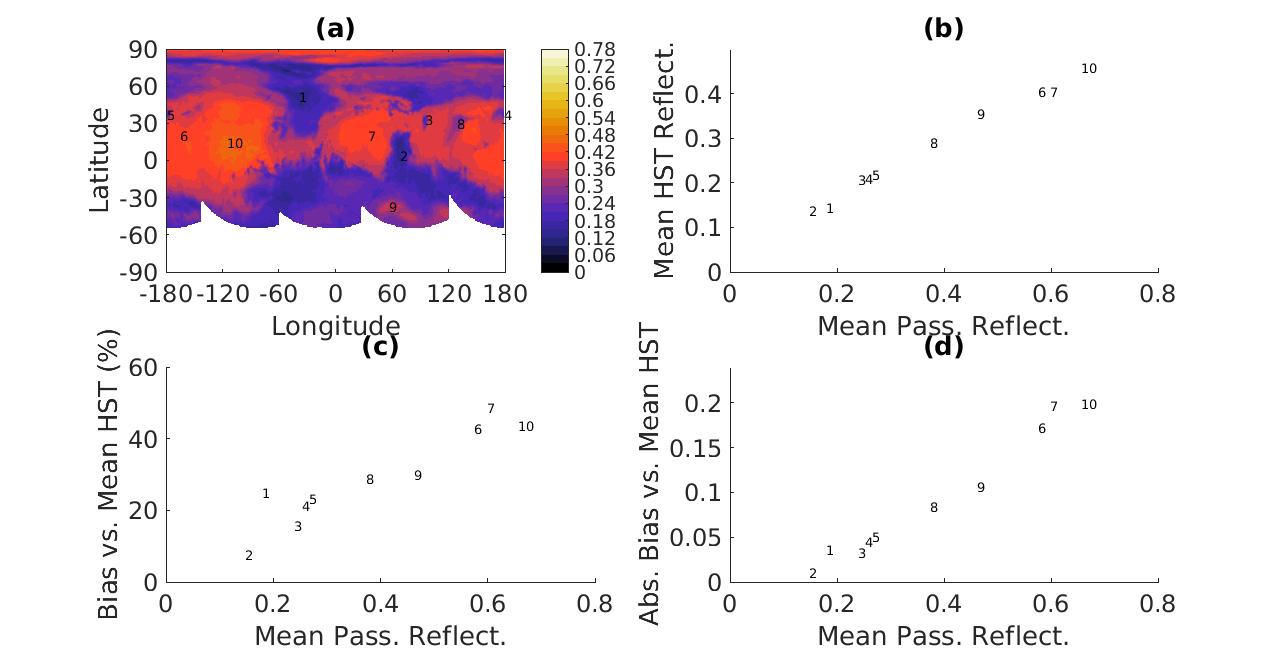}
\caption{Re-calibration analysis of the passive radiometry-based reflectivity map: (a) HST 1042 nm reflectivity map with the centers of the features used for re-calibration marked with numbers (exact boundaries are given in Table \ref{table:regionlist}; (b) Area-weighted mean passive radiometry-based reflectivity vs. area-weighted mean HST-based reflectivity; (c) Area-weighted mean passive radiometry-based reflectivity vs. bias relative to HST-based reflectivity (\%); (d) Area-weighted mean passive radiometry-based reflectivity vs. absolute bias relative to HST-based reflectivity.}  
\label{fig:recalib}
\end{figure}

Passive radiometry was re-calibrated by comparing spatial averages of the HST and passive radiometry-based maps in the centers of reflectivity features (Fig. \ref{fig:recalib}a). This analysis avoids the north pole. Focusing on the centers of features accounts for possible mismatches between the HST and passive radiometry data due to seasonal or secular reflectivity variability \citep{Szwast:2006}. Avoiding the north pole accounts for the possibility that HST observations underestimated reflectivity near the north pole (and any effects due to wavelength dependence of ice reflectivity). \citet{Sun:2006mola} notes that the passive radiometry data better resolves features near the pole than HST. Such features likely include the troughs of the north polar cap, which can be smaller than the 21.9 km maximum spatial resolution of the March 1997 observations. (These troughs are difficult to resolve at $1^{\circ}$ resolution in this analysis as well.) Indeed, the maximum reflectivity of the northern cap in the HST map is $\approx 0.48$, which is significantly less than the $> 0.6$ reflectance measured on the northern cap around northern summer solstice by \citet{Appere:2011ice}.

\begin{table}[ht]
\centering
\begin{tabular}{l l l}
\hline
\textbf{Region Number} & \textbf{Latitude Range} & \textbf{Longitude Range} \\
\hline
1 & $49^{\circ}$--$55^{\circ}$ N & $42^{\circ}$--$34^{\circ}$ W \\

2 & $10^{\circ}$ S--$18^{\circ}$ N & $64^{\circ}$--$74^{\circ}$ E \\

3 & $31^{\circ}$--$35^{\circ}$ N & $95^{\circ}$--$96^{\circ}$ E \\

4 & $36^{\circ}$--$39^{\circ}$ N & $179^{\circ}$--$180^{\circ}$ E \\

5 & $36^{\circ}$--$39^{\circ}$ N & $179^{\circ}$--$178^{\circ}$ W \\

6 & $10^{\circ}$--$30^{\circ}$ N & $179^{\circ}$--$150^{\circ}$ W \\

7 & $10^{\circ}$--$30^{\circ}$ N & $20^{\circ}$--$50^{\circ}$ E \\

8 & $28^{\circ}$--$32^{\circ}$ N & $125^{\circ}$--$133^{\circ}$ E \\

9 & $37^{\circ}$ S & $55^{\circ}$--$59^{\circ}$ E \\

10 & $10^{\circ}$ N & $120^{\circ}$--$110^{\circ}$ W \\

\hline
\end{tabular}
\caption{Regions used for the re-calibration of the passive radiometry data}
\label{table:regionlist}
\end{table}

The analysis confirms that there is an approximately linear monotonic relation between the passive radiometry-based and HST-based reflectivity estimates (Fig. \ref{fig:recalib}b). The relationship is only approximately linear. As expected from the regional analysis, lower reflectivity areas have smaller percentage biases than higher reflectivity areas (Fig. \ref{fig:recalib}c). This trend is quite non-linear. Therefore, the passive radiometry-based reflectivity was corrected by estimating the absolute bias in reflectivity between the HST map as a function of passive radiometry-based reflectivity. This data then could be interpolated or extrapolated to obtain a re-calibrated reference value ($R^{M,N,*}_{1064}$). In that case, the $R^{M,N,*}_{1064}$ for the peak value on the permanent water ice cap is 0.60.

\begin{figure}[ht]
\centering\includegraphics[width=1\linewidth]{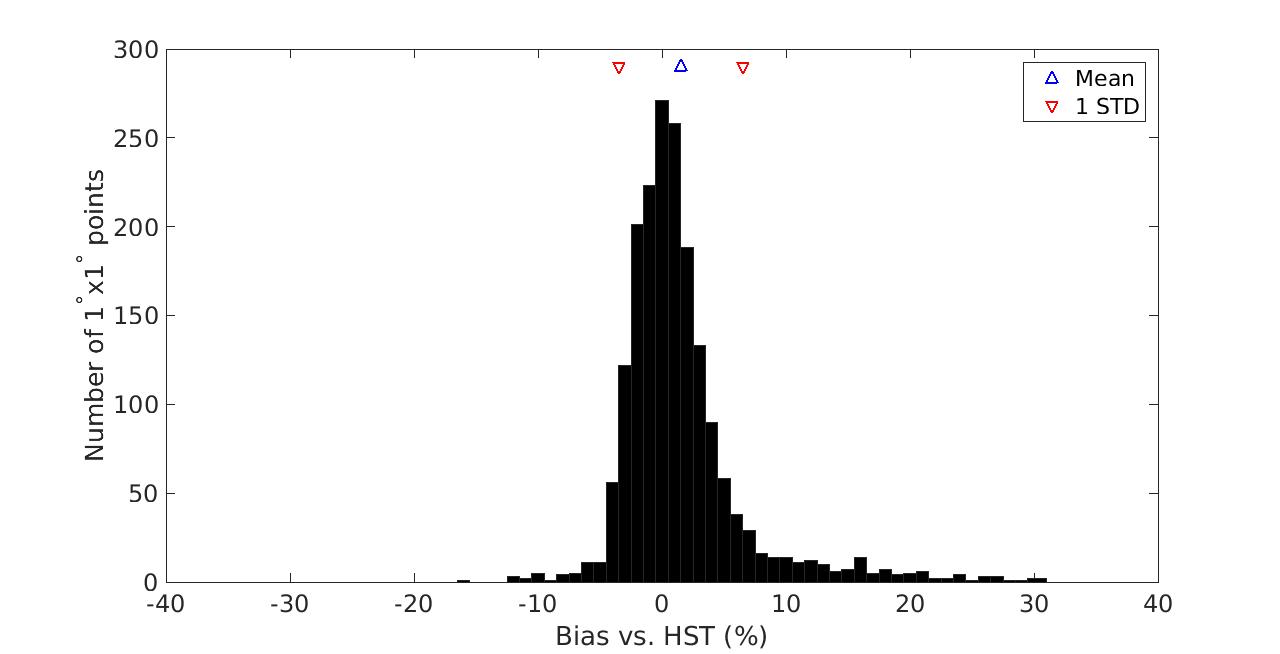}
\caption{Results of the re-calibration: (a) $R^{M,N,*}_{1064}$ mapped at $1^{\circ} \times 1^{\circ}$; (b) Frequency distribution of the bias of $R^{M,N,*}_{1064}$ vs. the HST-based reflectivity map at all $1^{\circ} \times 1^{\circ}$ points used for re-calibration}  
\label{fig:recalibmapanderror}
\end{figure}

The re-calibrated reflectivity data agrees well with the HST data (Fig. \ref{fig:recalibmapanderror}). There is a small, long tail of positive bias up to 30\%. However, the remainder of the bias distribution is centered at zero. The mean of the distribution is 1.5\%, and the median is 0.5\%. Including the long tail, $R^{M,N,*}_{1064}$ is within $+6.5\%/-3.5\%$ of the HST-based estimate ($\pm 1 \sigma$).  

\subsection{The Final Reference Product}
\begin{figure}[ht]
\centering\includegraphics[width=1\linewidth]{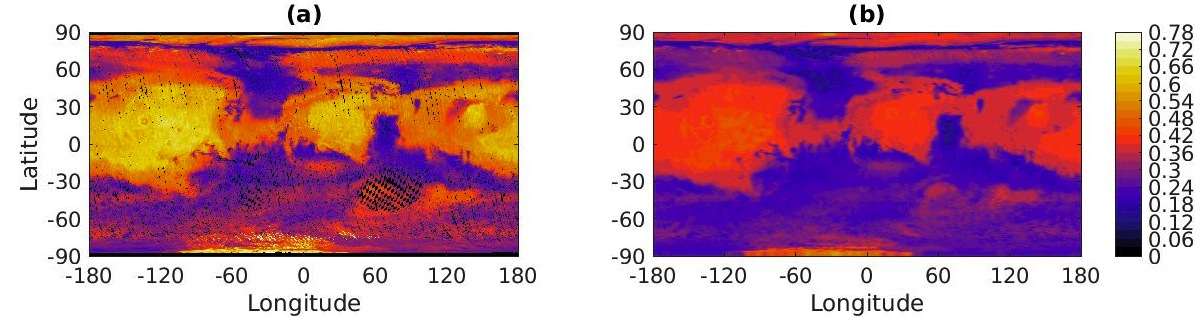}
\caption{(a) $R^{M,N}_{1064}$ from passive radiometry mapped at $0.5^{\circ} \times 0.5^{\circ}$. The areas in black indicate no suitable passive radiometry data was available to estimate $R^{M,N}_{1064}$; (b) Final reference product of $R^{M,N,*}_{1064}$ at $0.5^{\circ} \times 0.5^{\circ}$. See text for details.}  
\label{fig:finalrefproduct}
\end{figure}

The final reference product then can be generated. The product starts with a passive radiometry-based map of $R^{M,N}_{1064}$ at a resolution of $0.5^{\circ} \times 0.5^{\circ}$. Analysis at this resolution tests the resolution limits posed by the availability of suitable passive radiometry measurements. There is substantial missing data in the mid-latitudes, particularly in the southern hemisphere (Fig. \ref{fig:finalrefproduct}a). Moreover, a small number of bright points likely associated with the southern seasonal cap are resolved at higher southern latitudes. Finally, striped features are visible in the northern high latitudes and elsewhere.

Therefore, before re-calibration using the relations in Fig. \ref{fig:recalib}, values $>$ 0.50 between $82.5^{\circ}$--$65^{\circ}$ S were removed. Missing data was interpolated over along lines of constant latitude. Finally, the data was Fourier filtered along lines of constant latitude to remove high-frequency variability. Missing data at $> 87.5^{\circ}$ N and S were filled with the data at $87^{\circ}$ N and S. The error introduced by Fourier filtering averages 0.2\% with a standard deviation of 4.3\%, which is comparable to the bias of $R^{M,N,*}_{1064}$ relative to the HST-based reflectivity map. The final map is in Fig. \ref{fig:finalrefproduct}b and archived as \citet{Heavens:2016} along with the re-calibration data.

\subsection{Estimating the Reflectivity of the Seasonal Cap}

The final reference product provides $R^{M,N,*}_{1064}$ when the seasonal cap is not present. Estimating the reflectivity of the seasonal cap is more difficult. In Section \ref{S:2passrad}, it was described how this reflectivity could be estimated from passive radiometry observations during the period the cap was present. This idea takes advantage of the fact that the cap does not sublimate instantly when the area becomes sunlit.

\subsubsection{Passive Radiometry Estimates}

\begin{figure}[ht]
\centering\includegraphics[width=1\linewidth]{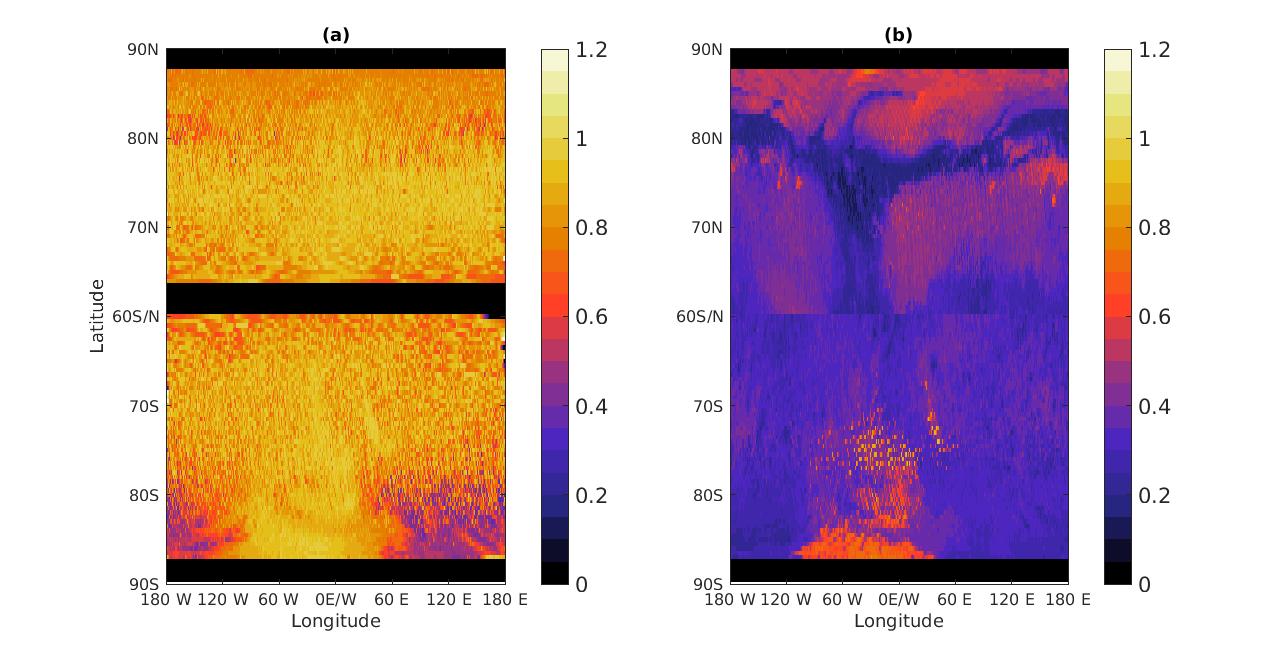}
\caption{(a) $R^{M,N}_{1064}$ of the seasonal cap from passive radiometry mapped at $0.5^{\circ} \times 0.5^{\circ}$. The areas in black indicate no suitable passive radiometry data was available to estimate $R^{M,N}_{1064}$; (b) $R^{M,N}_{1064}$ of the seasonal cap-free surface from passive radiometry mapped at $0.5^{\circ} \times 0.5^{\circ}$. Missing data was interpolated over along lines of constant latitude. The areas in black indicate no suitable passive radiometry data was available to estimate $R^{M,N}_{1064}$ at that latitude.}  
\label{fig:seasonalcap}
\end{figure}

The results of this estimate are compared with the standard estimate of $R^{M,N}_{1064}$ (when the cap is absent) in Figs. \ref{fig:seasonalcap}a-b. In the south, the brightest reflectivities (0.95--1) are observed on or near the permanent CO$_2$ cap (the bright area near the south pole in Fig. \ref{fig:seasonalcap}b), which extends from the south pole but is somewhat displaced to the west. Reflectivities of $\approx 0.9$ are observed at some longitudes at lower latitudes, but reflectivity can be as low as 0.6 in adjoining longitudes.

In the north, a similarly noisy area of 0.6--1 reflectivities is present from $60^{\circ}$--$80^{\circ}$ N. There are areas of 0.4--0.6 reflectivity in the dark areas on the far western and eastern margins of the permanent H$_2$O ice cap (visible as an area with reflectivity $> 0.5$ at the highest northern latitudes in Fig. \ref{fig:seasonalcap}b). Troughs in the cap that are visible in Fig. \ref{fig:seasonalcap}b can be seen in Fig. \ref{fig:seasonalcap}b also can be seen in Fig. \ref{fig:seasonalcap}a with careful stretching. The H$_2$O ice cap area itself has a seasonal cap reflectivity of 0.75--0.9 (Fig. \ref{fig:seasonalcap}a), one that is lower than the permanent cap area in the south.

\subsubsection{Interpretation and Re-calibration}
The estimates of seasonal cap reflectivity are complex and require some interpretation. The noisiness at lower latitudes somewhat traces the same kind of orbital-track features that imparts stripy texture to the MOLA-based reflectivity maps. As the seasonal cap reflectivity estimates were based on data from periods when the seasonal cap may not have been present (in order to keep the other reflectivity estimate from being contaminated by the seasonal cap), the variability here is likely due to some bins only containing data from times when the cap was present. The higher values of reflectivity thus better approximate the reflectivity when the seasonal cap was present.

Reflectivity is highest on and near the permanent southern CO$_2$ cap. The high reflectivity here is partly due to observing the seasonal cap where it is thickest, so there is a better chance of observing a "pure" CO$_2$ signature. A similar effect would be expected at the north pole, but the north pole is generally darker than the inferred seasonal cap at lower latitudes. 

This discrepant behavior between the areas of the two permanent caps is probably a result of the strong sensitivity of 1064 nm albedo to compositional/grain size variations in the first 10 $\mu$m of the surface. NIR observations of the caps show wide variations in the reflectivity of the polar caps with season that can be connected to deposition and/or removal of CO$_2$/H$_2$O ice/dust and/or grain size changes in the ices \citep{Bibring:2004spc, Langevin:2005npc, Appere:2011ice,Brown:2014spc}.

Therefore, the most practical approach to estimating the reflectivity of the seasonal cap when daylight passive reflectivity data is available is to use the raw reflectivity data from the nearest available passive radiometry data, convert it to normal reflectivity, and re-calibrate it in the same way as the standard reference product. Such an approach may incorporate the effects of clouds but will track the first order variability in surface reflectivity.

If the highest reflectivity on the permanent CO$_2$ cap (0.99) is estimated to be the approximate reflectivity of the "pure" seasonal cap, the re-calibration curve for the standard reference product implies that $R^{M,N,*}_{1064}$ of the seasonal cap is 0.78. This value is close to theoretical predictions by \citet{Bibring:2004spc} for the normal reflectivity of pure CO$_2$ ice at 1064 nm (0.80). However, note that cap reflectivity is quite sensitive to dust and water ice content as well as grain size \citep{Bibring:2004spc,Appere:2011ice}.  

\section{Meteorological Applications}
\label{S:4}

\subsection{Aerosol Column Opacity Estimates from MOLA Active Sounding Measurements}

The final reference product enables column opacity to be derived from Eq. \ref{eq:colopac}. Recall that comparison of this product with HST data suggests its $1 \sigma$ uncertainty is $\approx \pm 5\%$. The $1 \sigma$ uncertainty in $RT$ is likewise $\approx \pm 5\%$ \citep{Neumann:2003clouds}. Based on Eq. \ref{eq:uncert1}, the maximum possible $\delta\tau_{1064}$ for the estimated uncertainty bounds is $0.04$, which occurs when the errors are perfectly anti-correlated. The true uncertainty is likely smaller, but the maximum value is chosen to be conservative.

In many cases, it will be helpful to include saturated returns in averages (while accounting for uncertainty due to them). In that case, the estimated RT value for the saturated values is:

\begin{equation}
\label{eq:uncert2}
RT_{est}=\frac{1}{2}(RT_{sat}+R^{M,N,*}_{1064})
\end{equation}

Following the same approach as above in assuming the maximum error, $\delta\tau_{1064}$ is estimated to be:

\begin{equation}
\label{eq:uncert3}
\delta\tau_{1064}=\mp\frac{1}{2}log\bigg[(1.052)\frac{RT_{est}}{RT_{sat}}\bigg]
\end{equation}

However, note that Eq. \ref{eq:uncert2} is based on the principle that the best estimate of a number that falls in a given range is that it lies in the center of the range. In cases where unsaturated returns and saturated returns are close together in space, it is more likely that $RT_{est} \approx RT_{sat}$. Thus, estimates based on Eq. \ref{eq:uncert2} may be and will appear biased low on the margins and within gaps in opaque clouds. Accounting for this effect may be possible but is beyond the scope of this work. In the remainder of the paper, there will be references to the minimum detectable opacity threshold as a a result of the the detector saturation (and estimated by applying Eq. \ref{eq:colopac} to the case of: $RT_{est} = RT_{sat}$) as the sensitivity of the column opacity estimate.   

When taking spatiotemporal averages, there will be additional uncertainty due to variability over the domain of averaging. In that case, the uncertainty in the mean due to the uncertainty in the individual measurements will be treated as separate from the uncertainty due to the variance in the measurements ($\delta_{\tau_{1064}}^2$) in order to calculate a total uncertainty. The uncertainty in the mean ($\epsilon_{\tau_{1064}}$) then can be found by a standard square summation formulation:

\begin{equation}
\label{eq:uncert4}
\epsilon_{\tau_{1064}}^2=\frac{\sum_{k=1}^{n} \delta\tau_{1064}^2}{n^2}+\frac{\sigma_{\tau_{1064}}^2}{n}
\end{equation}

Fully absorbed returns are not returns \textit{per se}, so they will be not be considered when estimating column opacity. Their omission may introduce a negative bias into later analyses of mean opacity. Likewise out of consideration are non-surface returns, which do not provide sufficient information to evaluate column opacity without precise knowledge about aerosol backscattering.

Single wavelength lidar information alone cannot disambiguate dust from ice or one type of ice from another. However, a variety of instruments have demonstrated that there is a dominant aerosol component in particular locations and seasons on Mars. It is therefore valuable to apply a general aerosol column opacity to a variety of cases that are difficult to observe with other instruments.

In the remainder of this section, MOLA-based column opacity will be applied to various problems in Martian meteorology and climatology. The intention is not to fully test the climatological and/or meteorological hypothesis relevant to the example but instead to show how MOLA data could complement other datasets and techniques in performing such tests.

\subsection{Climatological/Meteorological Examples}

\subsubsection{Single Point Opacity Records}
\begin{figure}[ht]
\centering\includegraphics[width=1\linewidth]{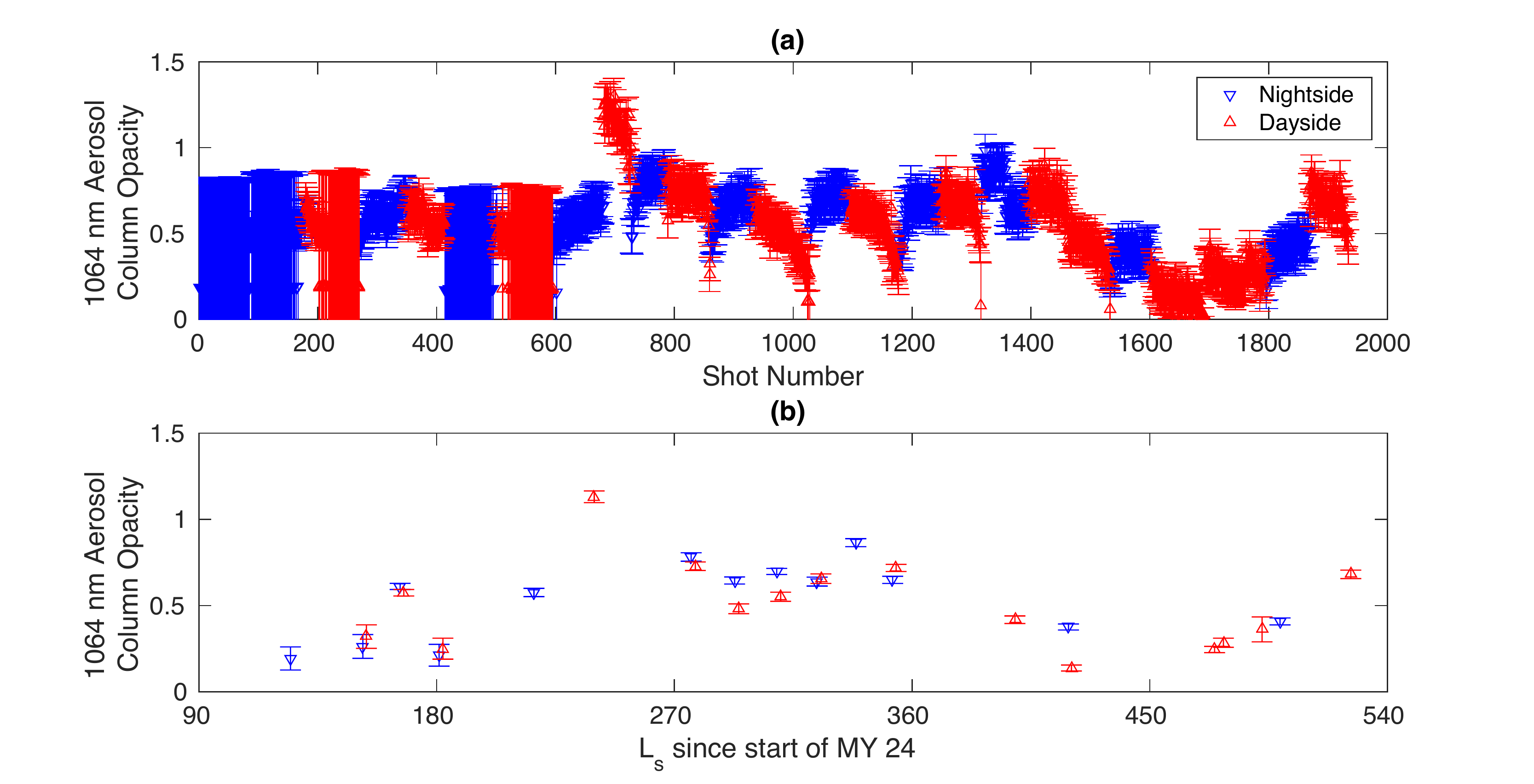}
\caption{Construction of an aerosol opacity record at $5.25^{\circ}$ S, $138^{\circ}$ E, an area within Gale Crater: (a) Raw column opacity measurements with 2$\delta$ single measurement error bars plotted in order of measurement; (b) Mean column opacity over the area during individual observational events plotted with 2$\epsilon$ uncertainty error bars. AM refers to 0:00--12:00 LST, PM to 12:00--24:00 LST. However, all observations here are near 2:00 and 14:00 LST.}  
\label{fig:colopac_dark}
\end{figure}

The simplest application of the MOLA active sounding data is to construct aerosol opacity records over small areas. The example shown is for a part of Gale Crater ($5.25^{\circ}$ S, $138^{\circ}$ E), the landing site of the Mars Science Laboratory (MSL) \textit{Curiosity} (Figs. \ref{fig:colopac_dark}a--b). This area is relatively dark ($R^{M,N,*}_{1064}=0.19$), so returns are mostly unsaturated after northern summer of MY 24. The seasonality and amplitude of this record agrees well with 880 nm opacity observations from MSL, which should be roughly equivalent to 1064 nm opacities based on diffuse scattering \citep{Montabone:2015opac, Guzewich:2016}.  

The peak mean opacity of 1.13 $\pm$ 0.02 occurred at $L_s=239.66^{\circ}$ of MY 24 (Fig. \ref{fig:colopac_dark}b). The individual measurements are as high as 1.4 (Fig. \ref{fig:colopac_dark}b). The origin of this opacity is somewhat mysterious. \citet{Cantor:2001ds} does not identify any local dust storms near Gale Crater at this time, and reports that a regional dust storm in the southern mid-latitudes had dissipated to a haze by $L_s=237^{\circ}$ of MY 24. 

\begin{figure}[ht]
\centering\includegraphics[width=1\linewidth]{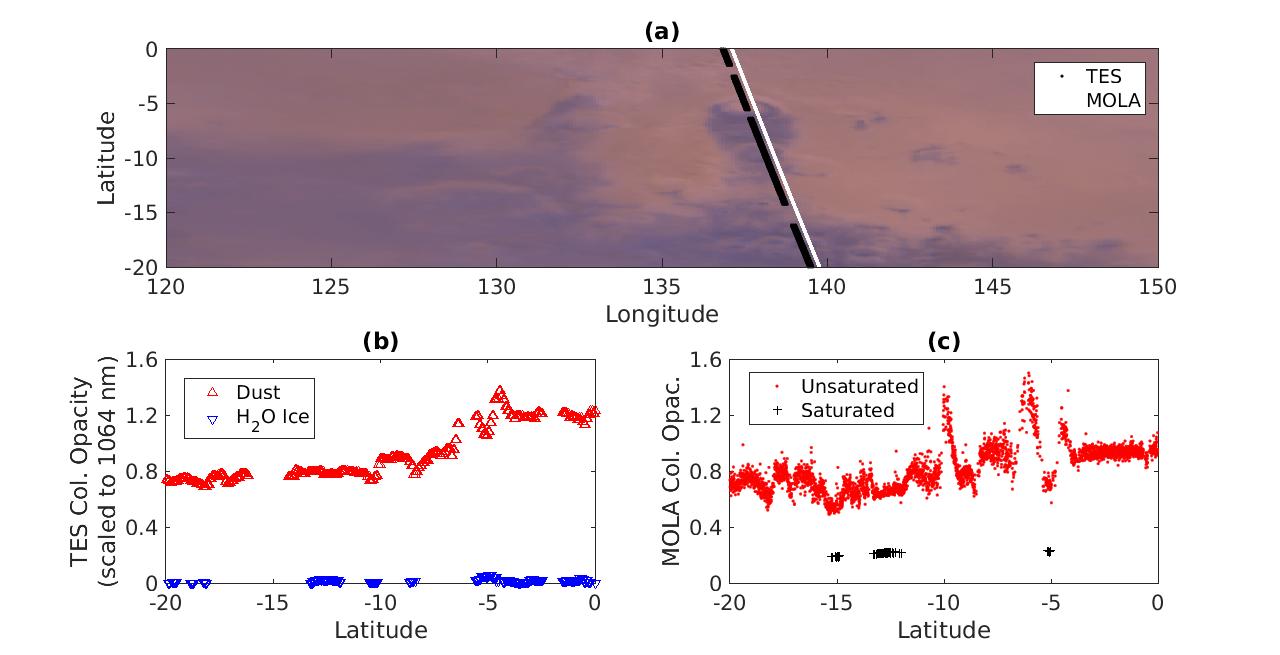}
\caption{(a) Mars Orbiter Camera image for the Martian sol centered at $L_s=239.67$ of MY 24. The Thermal Emission Spectrometer (TES) and MOLA tracks are plotted with black and white dots respectively; (b) TES retrieved column opacity for dust and water ice (scaled to MOLA) along the track in (a); (c) MOLA aerosol column opacity along the track in (b). Note that the column opacity is derived from all shots that appear to have surface returns, not just ones in Channel 1. The vast majority of returns, though, are in Channel 1.}  
\label{fig:tesmolaweirdopac}
\end{figure}

Additional constraints come from comparing MOLA data with TES retrievals and imagery from the Mars Orbiter Camera (MOC) on board MGS \citep{Wang:2014ds,Wang:2016}. The TES track is displaced $\approx$ 15 km to the west of the MOLA track but both pass over Gale and Lasswitz Craters ($5.5^{\circ}$ and $9.5^{\circ}$ S) (Fig. \ref{fig:tesmolaweirdopac}a). There are no obvious dust sources in the vicinity. TES retrievals suggest that dust is the dominant source of opacity (Fig. \ref{fig:tesmolaweirdopac}b). Dust opacity is $\approx 0.75$ from $20^{\circ}$--$10^{\circ}$ S and climbs to $\approx 1.2$ north of Gale and Lasswitz (Fig. \ref{fig:tesmolaweirdopac}b). Small peaks and troughs are associated with the craters themselves, but they are small enough to be attributable to mixing of dust into the deeper air column of the craters. 

The change in background opacity is somewhat muted in MOLA data. But MOLA resolves significant peaks in opacity associated with the craters, one in Lasswitz, and two on either side of Gale (Fig. \ref{fig:tesmolaweirdopac}c). The center of Gale is less dusty than its rim. (The small areas of saturated returns indicated areas where column opacity has fallen slightly below the background levels. The drop appears so large because of the conservative estimate that is made for the opacity of saturated returns.)  

It is thus possible that the craters themselves are the dust sources. The discrepancy between TES and MOLA observations may occur because TES observations are under-resolving the dust clouds, because the TES observations occur too far west to resolve them, or because the MOLA/TES opacity ratios are incorrect (because of atmospheric scattering or otherwise). One possible implication is that both craters were affected by small local dust storms ($< 60$ km diameter) on this particular sol. Information like this can be used to assess the accuracy and resolution limits of dust storm surveys based on weather camera data alone.

There are no pairs of overlapping dayside and nightside observations within the same sol. However, dayside and nightside observations in the same area do form pairs that occur within a few sols of one another. (These can be seen in nearly vertically overlapping points in Fig. \ref{fig:colopac_dark}b). The nightside opacity is significantly higher than the dayside opacity in three instances but lower in one instance.    

\subsubsection{The Aphelion Water Ice Cloud Belt}

\begin{figure}[ht]
\centering\includegraphics[width=1\linewidth]{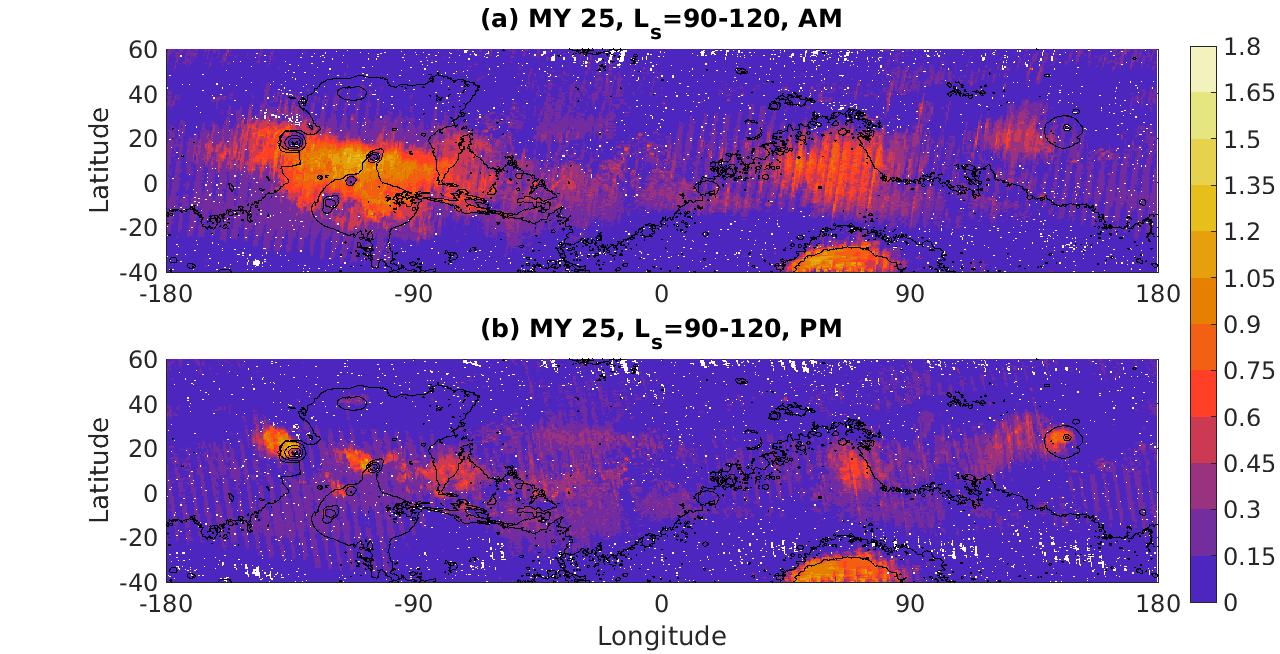}
\caption{Mean 1064 nm column opacity at $0.5^{\circ} \times 0.5^{\circ}$ resolution derived from MOLA Channel 1 surface returns for the labeled time intervals (color contours) and MOLA topography ($0.25^{\circ} \times 0.25^{\circ}$ resolution (5 km contours). Only values significantly greater than 0 (the difference of the mean and 2 $\epsilon > 0$) are plotted. Areas with no data are interpolated over on lines of constant latitude. AM refers to 0:00--12:00 LST, PM to 12:00--24:00 LST. However, all observations here are near 2:00 and 14:00 LST.}
\label{fig:watericecloudbelt}
\end{figure}

During portions of northern spring and summer, much of Mars's tropics are covered by water ice clouds of optical thickness sufficient to be seen in visible imagery \citep{WangandIngersoll:2002}. Moreover, opacity is dominated by water ice during this area and season \citep{Smith:2004tes}.  While the column opacity of these clouds can be evaluated on the dayside by visible and infrared observations, their column opacity on the nightside typically cannot be evaluated by either technique as a result of low light (for visible imagery) and poor thermal contrast between the cloud and the surface (for infrared nadir observations) \citep{Smith:2004tes}. 

It was realized by \citet{Wilson:2007waterice} that MOLA observations enabled the water ice cloud belt to be observed on the nightside for the first time. However, \citet{Wilson:2007waterice} does not provide much information about the methodology underlying the column opacity maps published therein. As best as can be inferred, the analysis of \citet{Wilson:2007waterice} uses the passive radiometry map of \citet{Sun:2006mola} to estimate reflectivity and mixes unsaturated and saturated returns.

For comparison, a mean column opacity map (using the methodology of this paper) for the same $L_s=90^{\circ}$--$120^{\circ}$ period in MY 25 analyzed by \citet{Wilson:2007waterice} is shown in Figs. \ref{fig:watericecloudbelt}a--b. These maps agree well with another in the areas with the highest water ice cloud opacities, such as the northwestern flank of Olympus Mons on the dayside. These maps also agree that the area of cloud cover expands significantly over Tharsis and Syrtis Major/E. Arabia on the nightside. Their disagreement primarily arises from the different approaches to handling saturated returns. 

The impact of sensitivity limits can be seen outside the region analyzed by \citet{Wilson:2007waterice}. The low reflectivity region of Syrtis Major ($8.4^{\circ}$ N, $69.5^{\circ}$ E) shows up as an area of thick cloud on the dayside. Analysis by \citet{Smith:2004tes} suggests that brighter areas to the east and west are similarly cloudy on the dayside, but only clouds over dark surfaces are detectable due to sensitivity limits.  

\subsubsection{Nightside Clouds Near Olympus Mons}
One notable feature of Figs. \ref{fig:watericecloudbelt}a--b is that clouds on the flanks of volcanoes like Olympus and Ascraeus Montes retreat to lower altitudes relative to the datum at night. As the sensitivity of the measurement over sub-seasonal timescales is primarily a function of the surface reflectivity, clouds closer to the summit likely have thinned in opacity. The expansion of clouds over the plateau areas near these volcanoes then can be interpeted as the thickening of clouds at lower altitudes. Taken together, both observations suggest that water ice condensation is occurring lower in the atmosphere.

Dayside water ice clouds over Olympus Mons and the Tharsis Montes tend to be thickest to the northwest of the summit \citep{Benson:2003waterice, Wilson:2007waterice} (Fig. \ref{fig:watericecloudbelt}b). However, it is unclear whether the nightside orientation is similar, though mesoscale modeling by \citet{Michaels:2006vwp} suggests that the distribution becomes more balanced between east and west at night.

\begin{figure}[ht]
\centering\includegraphics[width=1\linewidth]{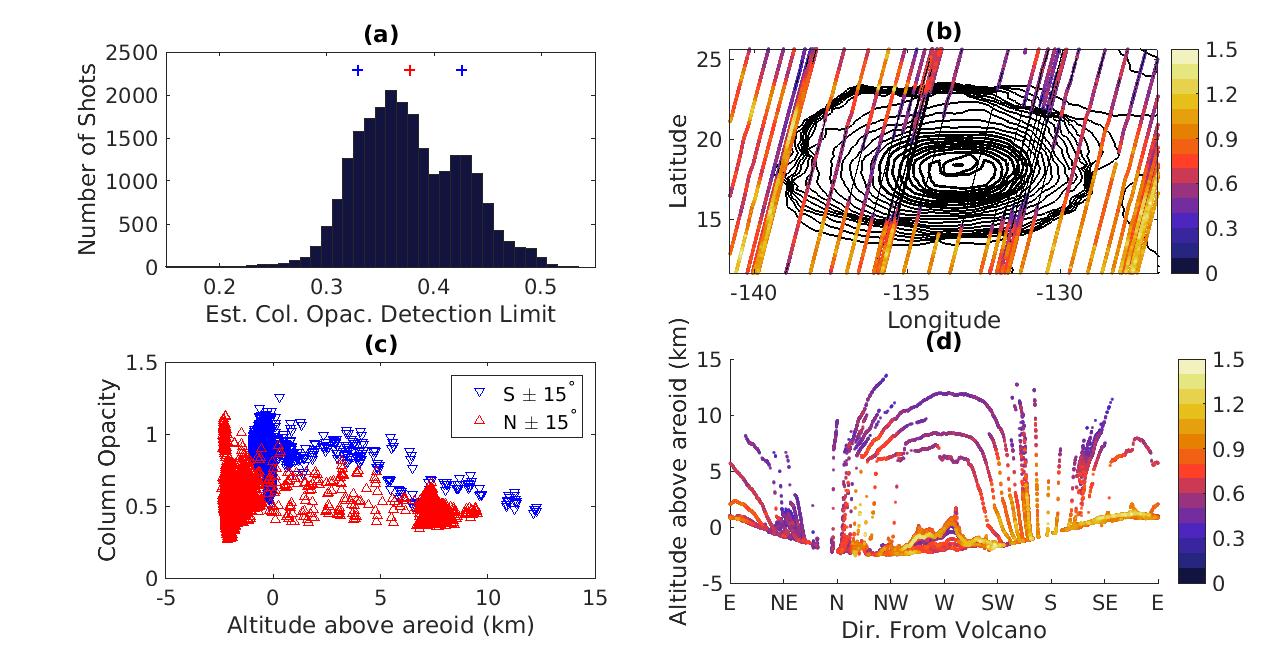}
\caption{Nightside clouds ($\approx$ 2:00 LST) over Olympus Mons ($11.65^{\circ}$--$25.65^{\circ}$ N, $135.8^{\circ}$--$121.8^{\circ}$ W) during $L_s=75^{\circ}$--$125^{\circ}$ of MY 25: (a) Histogram of the column opacity detection limit (the column opacity estimated from saturated surface returns in any channel and the local reflectivity) for this region and period. The mean and 1$\sigma$ uncertainties are indicated by the red and blue crosses; (b) Locations of individual unsaturated surface returns in any channel (colored dots) and saturated surface returns in any channel (black dots) plotted on MOLA topography (black contours). Dots are colored according to opacity; (c) Column opacity vs. altitude above the areoid for unsaturated surface returns in any channel sorted by direction relative to the summit of Olympus Mons, as labeled; (d) Locations of individual unsaturated surface returns in any channel (colored dots) plotted as a function of direction relative to the volcano and altitude above the areoid (km). Dots are colored according to opacity.}
\label{fig:olympusmonsnightcloud}
\end{figure}

Looking in the season during MY 25 when dayside clouds over Olympus Mons were greatest in extent \citep{Benson:2003waterice}, it can be shown that the sensitivity of the opacity estimate (the opacity implied by a saturated return) averaged $\approx$ 0.36 (Fig. \ref{fig:olympusmonsnightcloud}a). On the margins of the volcano, column opacities derived from unsaturated surface returns are rarely this low but decrease at higher altitudes above the areoid (Fig. \ref{fig:olympusmonsnightcloud}b). Thus, the closer a track comes to the summit, the longer the track of saturated returns. 

Looking at surface returns close to north or south of the summit is the one way to evaluate the trend of column opacity with surface elevation (one way to look at how the condensation level might vary with location). This analysis suggests that water ice clouds are generally thicker at any given surface elevation to the south of the summit (Fig. \ref{fig:olympusmonsnightcloud}c). Opacities fall below the sensitivity level at 9 km north of the summit and 12 km south of the summit. However, a fuller look at the relationship between opacity, altitude above the areoid, and direction from the summit suggests that measurable column opacity reaches surface elevations above the areoid of 13 km to the northwest of the summit (Fig. \ref{fig:olympusmonsnightcloud}d). In addition, measurable column opacities reach higher surface elevations to the west of the summit than to the east (Fig. \ref{fig:olympusmonsnightcloud}b and d). Therefore, considerable directional asymmetry remains in water ice cloudiness over Olympus Mons at night, though more careful comparison would be required to assess whether this asymmetry is large enough to be inconsistent with the simulations of \citet{Michaels:2006vwp}.          

\subsubsection{Nightside Clouds in Noctis Labyrinthus}
\begin{figure}[ht]
\centering\includegraphics[width=1\linewidth]{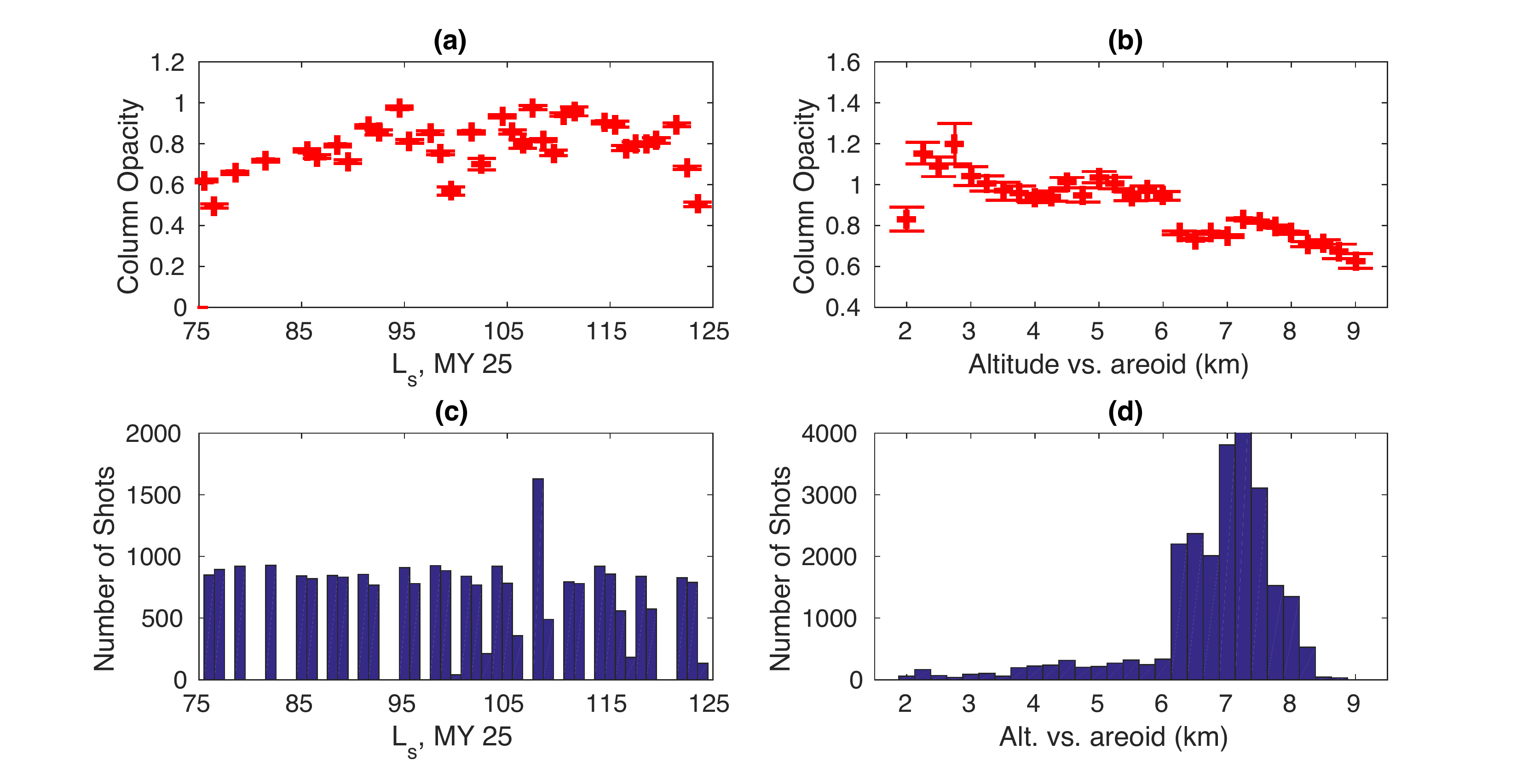}
\caption{Nightside clouds ($\approx$ 2:00 LST) near Noctis Labyrinthus ($10^{\circ}$--$4^{\circ}$ S, $104^{\circ}$--$94^{\circ}$ W) during $L_s=75^{\circ}$--$125^{\circ}$ of MY 25: (a) MOLA column opacity over the area estimated from surface returns in all channels as a function of $L_s$ (binning at $1^{\circ}$ of $L_s$). Error bars are $ \pm 2\epsilon$; (b) MOLA column opacity over the area estimated from surface returns in all channels as a function of surface elevation relative to the areoid (binning at 0.25 km). Error bars are  $\pm 2\epsilon$; (c) Histogram of surface returns in all channels vs. $L_s$ (binning at $1^{\circ}$ of $L_s$) (d) Histogram of surface returns in all channels vs. surface elevation relative to the areoid (binning at 0.25 km)}
\label{fig:noctislabyrinthusclouds}
\end{figure}

\begin{figure}[ht]
\centering\includegraphics[width=1\linewidth]{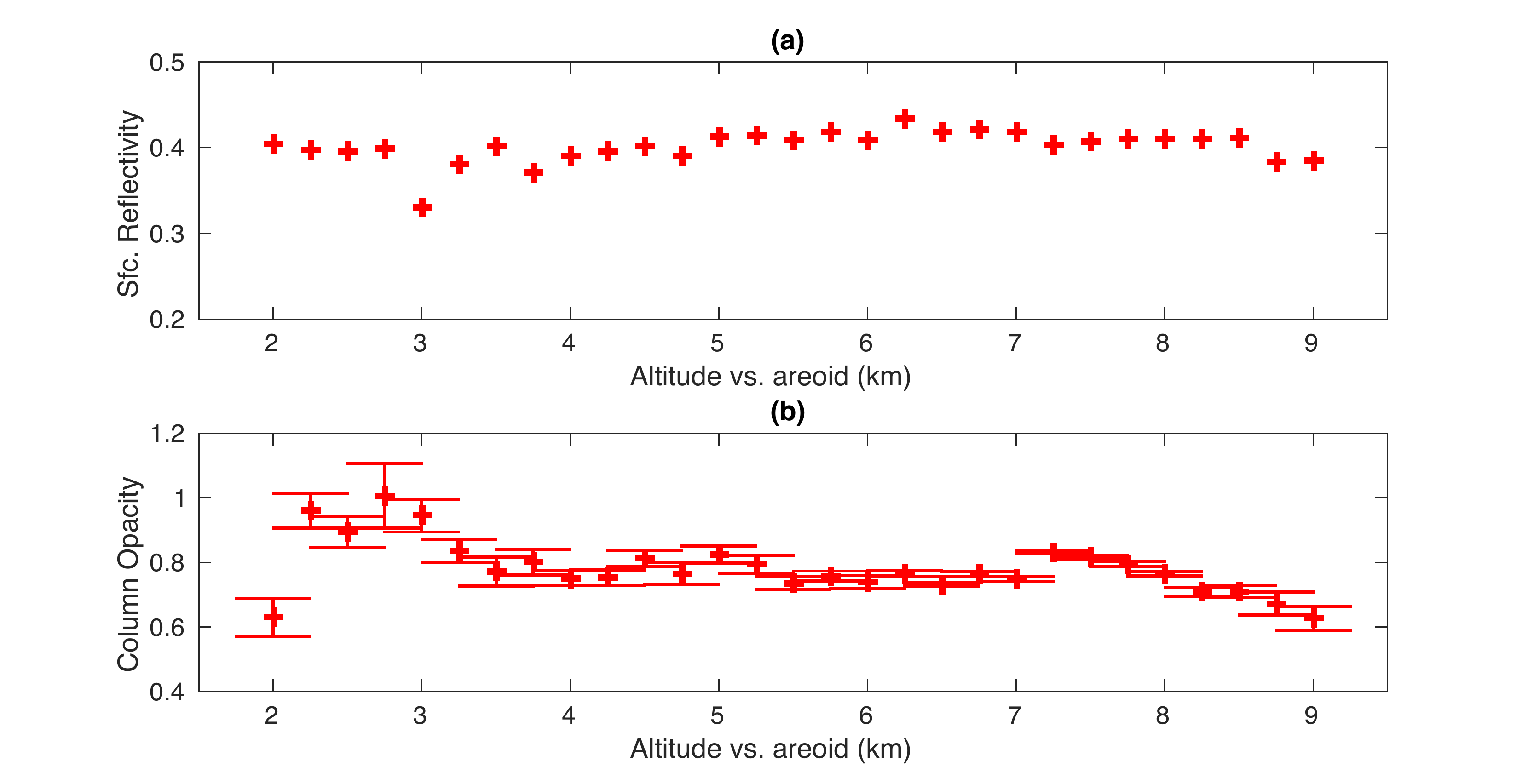}
\caption{Further analysis of nightside clouds ($\approx$ 2:00 LST) over near Noctis Labyrinthus ($10^{\circ}$--$4^{\circ}$ S, $104^{\circ}$--$94^{\circ}$ W) during $L_s=75^{\circ}$--$125^{\circ}$ of MY 25: (a) Mean surface reflectivity as a function of altitude relative to the areoid (binning at 0.25 km resolution) (b) MOLA column opacity over the area estimated from surface returns in all channels as a function of surface elevation relative to the areoid (binning at 0.25 km). Surface reflectivity at altitudes above the areoid less than 6 km is set equal to 0.27, a value typical of central Valles Marineris. Error bars are  $\pm 2\epsilon$.}
\label{fig:nlcorrect}
\end{figure}

Another feature of Fig. \ref{fig:watericecloudbelt}a is that the large area of nightside clouds around the Tharsis region covers the western portion of Valles Marineris known as Noctis Labyrinthus. As first noted by \citet{Briggs:1977}, clouds tend to concentrate in the canyon areas but not on the surrounding plateau. Mesoscale modeling, however, suggests that the canyon areas are generally warmer than the surrounding plateau, which implies that the canyons contain a water source that enables air within them to be saturated with respect to water ice when the plateaus are not \citep{Leung:2016vm}. In daylight hours, these clouds can be imaged and the contrasting cloud distribution between the plateau and canyons assessed. MOLA is the main option for observing these clouds at night.

Investigating in late northern spring and early northern summer (the season considered by \citet{Briggs:1977}), mean column opacities in Noctis Labyrinthus generally fall in the range of 0.5--1 (Fig. \ref{fig:noctislabyrinthusclouds}a). There is variability in mean column opacity throughout the period, particularly at the start, the end, and at $\approx L_s=100^{\circ}$; but there is little question that the area is consistently cloudy throughout the period. Note that the area is not consistently sampled at $1^{\circ}$ of $L_s$ cadence (Fig. \ref{fig:noctislabyrinthusclouds}c).

The histogram of topography shows that the area is dominated by a plateau at 6.5--8 km above the datum intersected by canyons at 2--6 km above the datum (Fig. \ref{fig:noctislabyrinthusclouds}d). The plateau has a column opacity during the period of $\approx 0.7$. At surface elevations of 2.25-6 km, mean column opacity is $\approx 1$ before dropping to 0.8 for surface elevations of 2 km. Therefore, it initially appears the canyons are $\approx$ 40\% cloudier than the plateau at night, except at the lowest altitudes.

However, the surface reflectivity product under-resolves the narrow canyons of Noctis Labyrinthus relative to areas further east in Valles Marineris (Fig. \ref{fig:finalrefproduct}b), which are significantly darker than the plateau surrounding Noctis Labyrinthus. Some contribution of the dark surfaces of the canyon to the reflectivity map are possible. Reflectivity is $\approx$ 0.40 at most altitudes but is lower in a few low altitude regions (Fig. \ref{fig:nlcorrect}a). If it is assumed that the canyon surfaces in Noctis Labyrinthus are as dark as typical canyons to the east in central Valles Marineris ($\approx$ 0.27), the canyons are as cloudy as the plateaus, except at altitudes relative to the areoid of 2.25--3 km (Fig. \ref{fig:nlcorrect}b). Note that these opacities are not normalized by altitude, so that altitude ranges where opacity is decreasing with altitude may imply the presence of clouds. In this case, cloud decks near 3 and 8 km above the areoid are plausible (Fig. \ref{fig:nlcorrect}b).   

\subsubsection{Mesoscale Structures in Dust Storms}

\begin{figure}[ht]
\centering\includegraphics[width=1\linewidth]{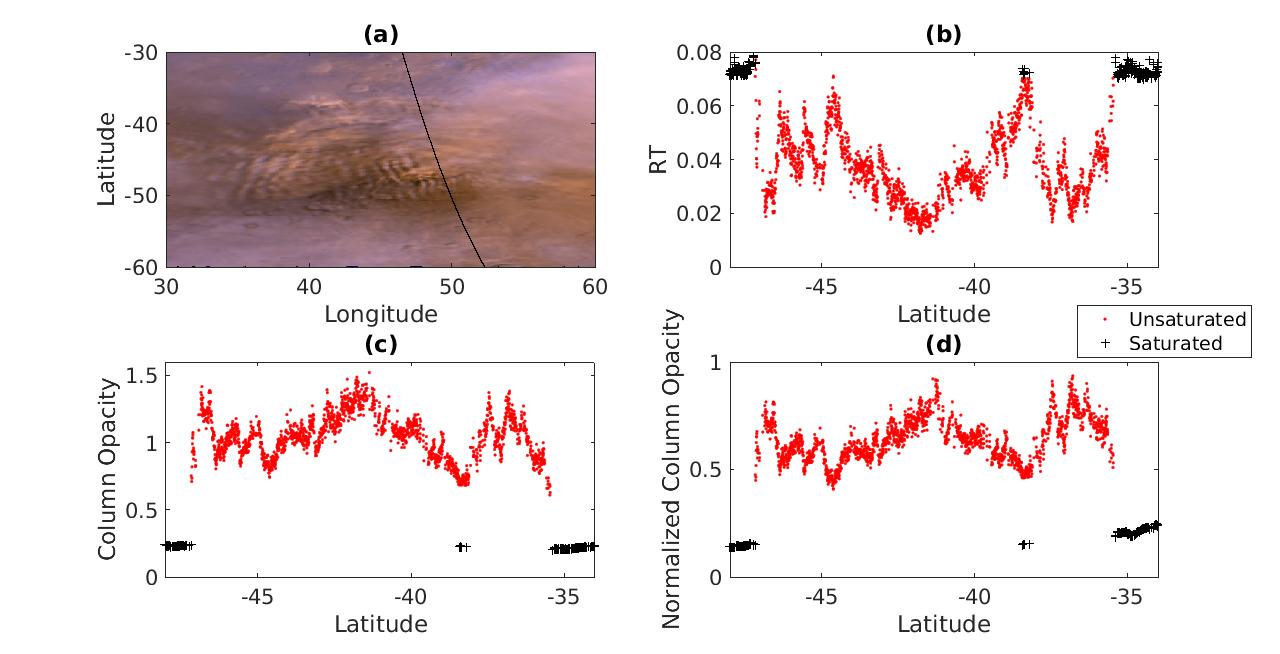}
\caption{Local dust storm at $L_s=151.42^{\circ}$ of MY 24 in the Hellas Basin: (a) Visible imagery from MOC (marsclimatecenter.com) \citep{Wang:2014ds}. The track of MOLA observations is plotted with black dots; (b) RT for surface returns in all channels in the vicinity of the storm; (c) 1064 nm column opacity for surface returns in all channels in the vicinity of the storm; (d) As in (c) but normalized by exp$(-z_{sfc}/H)$, where $z_{sfc}$ is the surface elevation relative to the datum and $H$ is a scale height of 10 km}
\label{fig:mesoscaledustdemo}
\end{figure}

Studies of Martian dust storms have mostly focused on synoptic to planetary aspects of dust storm structure, but mesoscale structure that is apparent in visible imagery is often described \citep{Briggs:1979dust,Cantor:2001ds, Strausberg:2005gds,Cantor:2007my25gds}. Typically, these descriptions contrast distinct turbulent features such as plumes and cells (on scales between the few km resolution of the imagery and $\sim$100 km) with indistinct laminar haze. \citet{Guzewich:2015texture} systematized this idea as "texture." \citet{Guzewich:2015texture} argues that texture indicates circulations associated with active dust lifting, because advection and diffusion would homogenize texture into hazes on timescales of hours. Due to the resemblance of some textures to convective clouds on the Earth, texture also has been interpreted as an indicator of convection as well as of active lifting \citep{Strausberg:2005gds}. Near-infrared imaging of a single local storm  has shown that optically thick dust storms can contain smaller, more opaque structures within them \citep{Maattanen:2009lds}. Yet structures in visible imagery usually are not interpreted in tandem with column opacity information.

MOLA can provide information about dust storm structure that either complements what can be inferred from visible imagery  or clarifies what is doubtful in it. The example shown in Fig. \ref{fig:mesoscaledustdemo} is of a local dust storm on the western margin of the Hellas Basin during late northern summer of MY 24 identified by the survey of \citet{Cantor:2001ds}. In the visible image, the MOLA track passes through an area of possible lee wave clouds that stretches well to the west (Fig. \ref{fig:mesoscaledustdemo}a). At $47^{\circ}$ S, $48^{\circ}$ E, these clouds are interrupted by a structures that are perpendicular but similar in wavelength to the possible lee wave clouds. The track then crosses a somewhat hazier area with longer wavelength structure from $46^{\circ}$--$38^{\circ}$ S. To the north of this area, the haze clears somewhat and finer wavelength structures are seen. 

The raw MOLA $RT$ data suggests that the wavy area on the southern margin of the storm is indeed associated with opacity fluctuations of $\approx$ 40 km wavelength and amplitude of $\approx$ 0.5 (Fig. \ref{fig:mesoscaledustdemo}b). Fluctuations with similar properties are also are observed on the northern margin of the storm. In the hazier central region of the storm, the dominant wavelength of variability increases to $\approx$ 500 km. Or viewed another way, opacity decreases from the center of the storm toward the wavy regions. Smaller wavelength structures may be embedded in this central area. Figs. \ref{fig:mesoscaledustdemo}c--d show that the structure in the RT data is due to neither reflectivity variations nor changes in the atmospheric column mass. The extent of the wavy regions (and the storm itself) cannot be assessed by MOLA due to the low sensitivity (high opacity threshold for unsaturated returns) implied by Figs. \ref{fig:mesoscaledustdemo}c-d. (This threshold is likely close to the lowest column opacity values reported for unsaturated returns of 0.7). Note that any additional radiance from aerosol forward scattering would mute the amplitude of variability.  

Storms on the western margin of Hellas are common at this season \citep{Cantor:2001ds}. It therefore would be possible to identify storms with similar presentation in visible imagery and analyze MOLA data over various parts of the storm to generate a composite structure for this storm type. Such an exercise is beyond the scope of this work but is currently underway for local dust storms in another area by the author. Note, however, that the storm analyzed here is an ideal case. In many cases, dust storms are so optically thick that they absorb the MOLA beam, which results in observational gaps and false cloud return artifacts \citep{Neumann:2003clouds}.

\section{Discussion}
\label{S:5}

\subsection{Improving the Reflectivity Map}
The final reference product presented in this paper was derived by a justifiable methodology but not necessarily a uniquely justifiable one. Improvement may be possible in four areas and could lead to reduced uncertainty relative to the validation dataset like the HST observations of \citet{Bell:1999hst}. However, progress in these areas would require substantial further work. In some cases, these efforts may not be worth the effort either because of the uncertainty introduced by secular reflectivity change or because of limitations in the validation dataset. 

First, the passive radiometry data that was not contemporaneous with active radiometry data was excluded. The excluded data is roughly five Earth years in duration and has six times better resolution than the passive radiometry contemporaneous with active radiometry \citep{Sun:2006mola}. This data could enable a higher resolution reflectivity product and/or improve results at the resolutions considered here by allowing tighter constraints on phase angle and/or opacity etc.

However, incorporating this data presents challenges. Cloud filtering by TES is possible, but the centers of TES retrievals are typically displaced by 15 km or so from MOLA, and as discussed in Section \ref{S:4}.1 may not resolve some cloud features resolved by MOLA. Furthermore, the higher the resolution of observations, the more likely it is that they will resolve features not resolved at the 21.9 km maximum resolution of the HST data used to create the recalibration dataset \citep{Bell:1999hst}. Moreover, using more data from beyond the period of active radiometry increases the probability of bias due to secular reflectivity change.

Secular reflectivity change is likely not as important on decadal timescales as once was thought but still might be significant in the particular case considered here. \citet{Bell:1999hst}, for example, estimated secular change over 20 years of $\pm$ 25\%, However, \citet{Szwast:2006} demonstrated with higher temporal resolution data from TES that most reflectivity changes due to large-scale dust storm activity last 1--2 Mars Years in duration and return to pre-storm levels thereafter. Yet adding passive radiometry data after the end of active radiometry data would add almost 3 Mars Years after the 2001 global dust storm (and the attendant secular variability). To a lesser extent, secular variability in reflectivity may affect all comparisons between HST and MOLA-based reflectivity maps. The HST dataset itself may not be an accurate measurement of the reflectivity during MOLA active sounding.

Second, the photometric model could be changed to incorporate insights from multi-angle observations by NIR observations subsequent to MOLA \citep{ Esposito:2006pfs,Vincendon:2007omega}. Any such changes would require re-analysis of the HST observations to match the new photometric model. However, as the differences between HST-based and MOLA-based reflectivity estimates seem mostly due to a bias in MOLA passive radiometry, it is unclear whether improving photometry would reduce uncertainty.

Third, the source of the offset between HST-based and MOLA-based reflectivity estimates could be determined and MOLA reflectivity measurements re-calibrated to account for the source of the difference. However, this offset does not have an obvious origin and is somewhat non-linear in reflectivity (Figs. \ref{fig:recalib}b--d).   

Fourth, the seasonal cap could be distinguished by using the raw seasonal cap extent data in the supplemental data of \citep{Piqueux:2015cap}. This improvement would be the easiest to implement of those considered but was omitted here for the sake of expediency.

\subsection{MOLA Data and Future Lidar Projects}
In recent years, there has been growing interest in sending an orbital lidar instrument to Mars \citep{Singh:2011lidar, Abshire:2015lidar, Brown:2015lidar}. The results presented here and the applications that follow from them can support justifying and planning for such an instrument in three crucial ways.

First, MOLA data illustrates how widespread mesoscale structures are in Mars's atmosphere and thus underlines why horizontal resolution matters when deciding between lidars and other types of instruments when pursuing certain measurement objectives. For example, some proposed Mars orbital lidars emphasize their capability to measure Martian winds \citep{Singh:2011lidar, Abshire:2015lidar}. Yet winds also can be measured by sub-mm/microwave sounders, which have estimated horizontal resolutions at nadir viewing of 30 km \citep{Kasai:2012submm}. Proposed lidars have similar wind measurement precision at this resolution \citep{Abshire:2015lidar}. 

The advantages of lidar are easier to see when considering measuring water vapor, which the proposed instrument of \citet{Brown:2015lidar} could measure at better than 100 m resolution, two orders of magnitude better than a sub-mm sounder \citep{Kasai:2012submm}. But there is little point in doing so if variability in the water vapor distribution is dominated by scales much longer than 30 km. MOLA data, however, suggests that the column aerosol distribution can vary significantly on scales of 30 km or less. There are 40 km wavelength waves on the margins of the western Hellas dust storm (Fig. \ref{fig:mesoscaledustdemo}), possible crater-sized local dust storms in the southern tropics (Fig. \ref{fig:tesmolaweirdopac}), and evidence for thicker nocturnal water ice clouds in a particular altitude range within the 10 km wide canyons of Noctis Labyrinthus (Figs. \ref{fig:noctislabyrinthusclouds} and \ref{fig:nlcorrect}). And if aerosol structures vary on these scales, water vapor likely varies similarly as well.         

Second, MOLA data can help better define the measurement and data processing requirements for future lidar systems, as \citet{Brown:2015lidar} is aware. The results of this study underline the importance of reflectivity in designing future lidar products. Algorithms to derive information from lidar observations will need good initial guesses for reflectivity as well as a way of adjusting those guesses in response to the observations. Lidar measurements near nadir may need to correct for opposition effects and aerosol scattering. These corrections likely will require dedicated campaigns of multi-angle observations by the lidar itself. Another important point is how important validation is. If it were not for careful forethought by \citet{Bell:1999hst}, it would be difficult to identify the biases in active sounding and passive radiometry.

Third, MOLA data can be a testbed for testing science integration and science closure strategies for orbital science payloads that include lidars. As shown in examples like Fig. \ref{fig:tesmolaweirdopac}, data from visible imagers, infrared sounders, and lidars are most powerful when interpreted together. Furthermore, as the Olympus Mons and Noctis Labyrinthus nightside cloud examples suggest, the high horizontal resolution of lidar systems makes them ideal for testing hypotheses based on mesoscale atmospheric models and posing new problems for those models to attack.

\section{Summary}
\label{S:6}

This study has analyzed and mapped the 1064 nm reflectivity of Mars with MOLA data in order to derive a 1064 nm aerosol column opacity product from surface returns measured by MOLA active sounding. The main conclusions are that: (1) the MOLA detector saturation problem and uncertainties relating to surface and aerosol scattering near zero phase angle makes it impossible to derive reflectivity from active sounding measurements over much of Mars's surface; (2) the differences between \citet{Sun:2006mola}'s analysis of 1064 nm reflectivity and HST-based estimates of 1042 nm reflectivity \citep{Bell:1999hst} are likely due to a significant, non-linear bias in MOLA passive radiometry measurements rather than contamination by clouds or differences in observational geometry; (3) the resulting column opacity product can improve studies of climatology and mesoscale meteorology on Mars while enabling better justification of and planning for future orbital lidars at Mars.

\section*{Acknowledgments}
This work was supported by NASA's Mars Data Analysis and Solar System Workings Programs (NNX14AM32G; NNX15AI33G). The author thanks Jim Bell and Mike Wolff for providing him with the Hubble Space Telescope-based 1042 nm reflectivity data. He also thanks John Wilson and an anonymous reviewer for their extremely helpful reviews.



%% References

\section*{References}
\bibliographystyle{elsarticle-harv}
\bibliography{molabrightart_1.bib}






% \end{thebibliography}

\end{document}